\documentclass[manuscript]{aastex}

\slugcomment{Revised and submitted to The Astrophysical Journal}

\shorttitle{Late Formation of Saturn's Moons}
\shortauthors{\' Cuk, Dones \& Nesvorn\' y }

\begin{document}

\title{Dynamical Evidence for a Late Formation of Saturn's Moons}

\author{Matija \' Cuk}
\affil{Carl Sagan Center, SETI Institute, \\
 189 North Bernardo Avenue, Mountain View, CA 94043\\
\email{mcuk@seti.org}}

\author{Luke Dones}
\affil{Southwest Research Institute, \\
%1050 Walnut St., Suite 400, 
Boulder, CO 80302}

\and

\author{David Nesvorn\' y}
\affil{Southwest Research Institute, \\
%1050 Walnut St., Suite 400, 
Boulder, CO 80302}

\begin{abstract}
We explore the past evolution of Saturn's moons using direct numerical integrations. We find that the past Tethys-Dione 3:2 orbital resonance predicted in standard models likely did not occur, implying that the system is less evolved than previously thought. On the other hand, the orbital inclinations of Tethys, Dione and Rhea suggest that the system did cross the Dione-Rhea 5:3 resonance, which is closely followed by a Tethys-Dione secular resonance. A clear implication is that either the moons are significantly younger than the planet, or that their tidal evolution must be extremely slow ($Q > 80,000$). As an extremely slow-evolving system is incompatible with intense tidal heating of Enceladus, we conclude that the moons interior to Titan are not primordial, and we present a plausible scenario for the system's recent formation. We propose that the mid-sized moons re-accreted from a disk about 100 Myr ago, during which time Titan acquired its significant orbital eccentricity. We speculate that this disk has formed through orbital instability and massive collisions involving the previous generation of Saturn's mid-sized moons. We identify the solar evection resonance perturbing a pair of mid-sized moons as the most likely trigger of such an instability. This scenario implies that most craters on the moons interior to Titan must have been formed by planetocentric impactors.
\end{abstract}


\section{Introduction}

One of the most important results of the Cassini mission to Saturn was the discovery of a high heat flow from the south polar region of Enceladus \citep{por06}. This heat flow is estimated at 10-15 GW \citep{how11, how14}, which is by an order of magnitude in excess of theoretical predictions assuming a dynamically steady state \citep{mey07}. A dynamical equilibrium means that Enceladus maintains a constant orbital eccentricity, which corresponds to an equilibrium between excitation by the 2:1 orbital resonance with Dione and damping by tidal dissipation within Enceladus. When calculating this dynamical equilibrium, \citet{mey07} assumed that the strength of tidal dissipation within Saturn (which pushes Enceladus into the orbital resonance) can be quantified by a tidal quality factor $Q=18,000$, as suggested by \citet{sin83}. Tidal $Q=18,000$ corresponds to the strongest tidal response by Saturn that is still consistent with Mimas not evolving past its current location over 4.5 Gyr \citep{sin83, md99}. Any larger tidal evolution rates (corresponding to smaller tidal Qs) would place Mimas within Saturn's rings less than 4.5 Gyr ago. 

\begin{table}[!ht]
\begin{center}
\caption{Masses and mean orbital parameters of Saturn's major moons and their Trojan companions (semimajor axes of Trojans are marked by superscript ``t''). All quantities were retrieved from the Jet Propulsion Laboratory's Solar System Dynamics website ssd.jpl.nasa.gov on 04/13/15. Semimajor axes are scaled using Saturn's equatorial radius of 60,268~km, and inclinations are measured relative to the relevant Laplace planes. Masses of the Trojan moons (in italics) are estimates based on an assumed density of 500~kg~m$^{-3}$. Eccentricities and inclinations shown here are ``free'', and do not include the forced components \citep[see ][]{md99}. In the cases of Enceladus and Hyperion, typical instantaneous eccentricities ($e=0.005$ for Enceladus and $e=0.1$ for Hyperion) are well in the excess of the free component, so the eccentricity shown is starred.\label{elem}}
\bigskip
\begin{tabular}{|l|c|c|c|c|}
\hline\hline
Moon & Mass & Semimajor & Eccentricity & Inclination \\ 
\ & ($10^{20}$~kg) & axis ($R_S$) & & ($^{\circ}$)\\ 
\hline\hline
Mimas & 0.375 & 3.079 & 0.0196 & 1.574\\
\hline
Enceladus & 1.08 & 3.950 & 0.0000$^*$ & 0.003\\
\hline
Tethys & 6.176 & 4.889 & 0.0001 & 1.091 \\
\hline
Calypso & ${\it 2.5 \times 10^{-5}}$ & 4.890$^t$ & 0.0005 & 1.500 \\
\hline
Telesto & ${\it 4.0 \times 10^{-5}}$ & 4.890$^t$ & 0.0002 & 1.180 \\
\hline
Dione & 10.96 & 6.262 & 0.0022 & 0.028\\
\hline
Helene & ${\it 1.1 \times 10^{-4}}$ & 6.262$^t$ & 0.0000 & 0.213\\
\hline
Polydeuces & ${\it 4.5 \times 10^{-8}}$ & 6.259$^t$ & 0.0191 & 0.175\\
\hline
Rhea & 23.07 & 8.745 & 0.0002 & 0.333\\
\hline
Titan & 1345 & 20.27 & 0.0288 & 0.306\\
\hline
Hyperion & 0.059 & 24.90 & 0.0232$^*$ & 0.615\\
\hline
Iapetus & 18.06 & 59.08 & 0.0293 & 8.298\\
\hline
\end{tabular}
\end{center}
\end{table}

There are other possible approaches to constraining the tidal response of Saturn. One is direct observation of the moons' tidal acceleration. Secular expansion of satellite orbits with orbital periods longer than the planet's spin period is predicted by tidal theories \citep{md99} but was previously thought to be too slow for detection. Recently, \citet{lai12} re-analyzed a large number of historical observations of Saturn's moons and found tidal recession corresponding to $Q=1700\pm500$. More recent preliminary results using Cassini-based and ground-based observations of the satellites apparently support the idea of such fast tidal evolution of Saturn's moons \citep{lai15}. \citet{lai12} also point out that $Q\simeq1700$ matches what is required for Enceladus to be in tidal equilibrium and still produce the observed heat flux. However, such fast tidal rates would require the inner moons like Mimas to have formed less than about 0.5 Gyr ago (as they would have otherwise migrated past their present positions, assuming the $Q$ of Saturn did not change over time).  Theoretical models of Saturn's interior dissipation are also available, but they are not conclusive to this debate, as there is a basic disagreement about whether the dissipation happens in the core or the envelope \citep{ogi14, rem15, sto15}.

We can also constrain Saturn's tidal response by theoretical modeling of present and past orbital resonances between different moons. Low Q (i.e. a high rate of dissipation) of Saturn on the order of $Q=1700$ would make the apparent highly-evolved state of the Titan-Hyperion resonance a natural product of tidal evolution over the age of the Solar System \citep{gre73, cuk13}, whereas in the classical picture the resonance had to be primordial, as it would not have not changed significantly over 4.5 Gyr \citep{pea99}. Here we address the two most important orbital resonances that have likely happened in the past: the 3:2 mean-motion resonance (MMR) between Tethys and Dione, and the 5:3 MMR between Dione and Rhea. These resonances should have happened close in time about 1.1 Gyr ago if $Q=18,000$ \citep{md99, mey07}, or about 100 Myr ago if $Q=1700$. The 3:2 Tethys-Dione resonance has previously been studied using semi-analytical methods, which found substantial excitation of the two moons' orbital eccentricities, but ignored orbital inclinations \citep{zha12}. This likely eccentricity excitation was also suggested to be related to geological features on Tethys that indicate past tidal heating \citep{che08}. Unlike eccentricities, which damp over time by tides, the moons' orbital inclinations cannot be easily erased once they are excited \citep{che14}. Therefore, it is essential to probe the effects of possible past resonances on the moons' inclinations, using methods that account for the full dynamics of the system.

\section{Numerical Methods}

In this work we use two different numerical integrators, {\sc simpl} and {\sc complex}. {\sc Simpl} is relatively efficient, but cannot model close approaches between massive bodies and is applicable only to the situations where the moons' orbits are well-separated. {\sc Complex} is capable of integrating close encounters between moons, but has a downside of being much slower than {\sc simpl}.

With a goal of studying the long-term tidal evolution of Saturn's moons, we wrote a new symplectic integrator, {\sc simpl} (Symplectic Integrator for Moons and PLanets). This is a combination of two Levison-Duncan-type integrators \citep{lev94}, one for the planets and one for the moons of one planet. This type of integrator is a ``democratic heliocentric" (i.e. uses heliocentric coordinates and does not require strict ordering of planets) version of the mixed-variable symplectic integrator of \citet{wis91}, which assumes bodies follow Keplerian orbits and implements perturbations as discrete ``kicks" executed in Cartesian coordinates. We adapted the specific algorithm from \citet{cha02}, who designed it to model the dynamics of planets around one component of a binary star. {\sc simpl} directly includes mutual moon-moon interactions, as well as solar and planetary perturbations (direct planetary perturbations can be switched off to speed up computations; back reaction of satellites on planets is ignored). We also include the planet's oblateness, tides raised on the planet by the satellite (as a simple tangential acceleration on the moon in the plane of its orbit, not changing the planet's rotation) and satellite tides (implemented as radial-only kicks), and artificial planetary or satellite migration (to approximate interaction with protoplanetary or protosatellite disks). When calculating tidal accelerations, we assumed that the tidal quality factor $Q$ of Saturn was independent of the satellite's orbital frequency (i.e. all the moons experienced the same $Q$ for Saturn). We extensively validated {\sc simpl} against published results and analytical estimates, and found that all precessional motions and secular tidal effects are reproduced correctly.

To integrate the orbits of moons like Tethys and Dione with orbital periods of 2-3 days, we used a 0.1 day ($2.5 \times 10^{-4}$~yr) integration time-step. This is the timestep we used in most simulations shown in this paper, unless otherwise noted. We always included Titan in our simulations, and we usually included only Jupiter and Saturn in the heliocentric part of the integration. 

{\sc complex} (Crossing Orbit Moon and PLanet EXtrapolator) is a more basic type of symplectic integrator, in which the Hamiltonian is not separated into Keplerian and perturbation parts (as in {\sc simpl}) but into kinetic and potential energy. The exact implementation has been taken from \citet{for90}, and uses a fourth order algorithm. {\sc complex} uses many of the same subroutines as {\sc simpl} and the same input and output files, except that {\sc simpl}'s Keplerian step has been replaced by a sequence of symplectic mappings in Cartesian coordinates. The subroutines in {\sc simpl} that implement perturbations from other bodies, the planet's figure and tidal deformations have been incorporated into the symplectic mapping, in the sub-step where the momenta are being modified due to potential energy. 

{\sc complex} has no preference for Keplerian orbits, and can therefore resolve close approaches between the moons naturally. The downside of {\sc complex} is  its much slower speed, requiring about 600 timesteps for the shortest orbital period. For the sake of simplicity, we decided to use the same routines to integrate the satellites and the planets in {\sc complex}, although the planetary orbits could in principle be done more efficiently. Overall, {\sc complex} is about two orders of magnitude slower than {\sc simpl}, and we used it only when the satellite system went through an instability. Short collision times between moons (centuries to millennia) keep the relative inefficiency of {\sc complex} from becoming a major issue. The switching between {\sc simpl} and {\sc complex} was done manually, meaning that we go back to the state of a {\sc simpl} simulation just before the instability and continue the simulation using {\sc complex}.  

While the timestep of {\sc complex} was sufficiently small to resolve close approaches, it was still too large to reliably resolve collisions. Therefore we shrunk the timestep by an additional factor of 20 when two moons were between 10 combined radii from each other. Once again, relatively large collisional cross sections of the moons compared to the sizes of their orbits made this relatively inefficient method workable. Note that this approach would not work for simulations of planets interacting with small bodies, where special treatment for the two bodies undergoing the close encounter \citep{lev94} makes more sense than slowing down the whole integration.

\section{Tethys-Dione 3:2 Resonance}

\begin{figure}
\epsscale{0.8}
\plotone{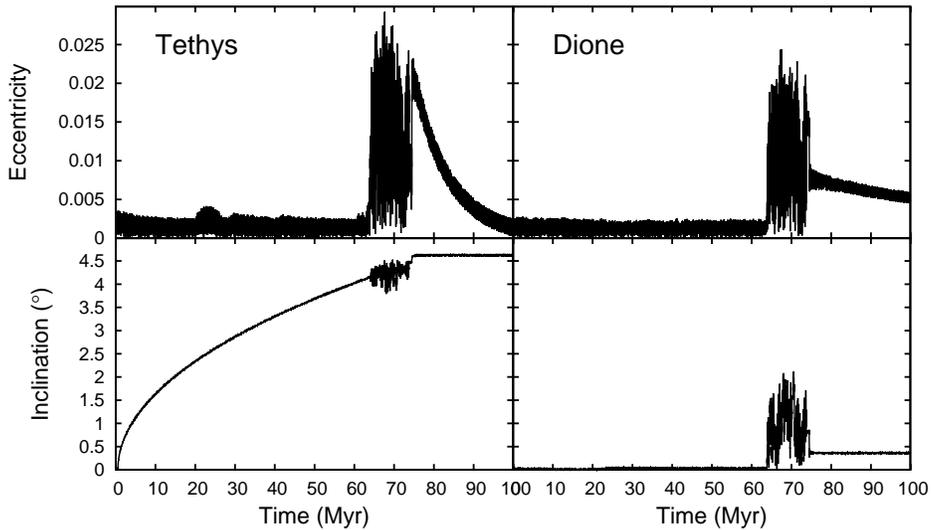}  
\caption{A simulation of Tethys and Dione encountering their mutual 3:2 resonance, using our integrator {\sc simpl}. The tidal response of Saturn was $Q/k_2=4000$, while the satellites were assumed to have tidal parameters $Q=100$ and $k_2=0.05$. We assumed that both moons had low eccentricities and inclinations before the resonance occurred. At first the moons are captured in the $i_1^2$ 4:2 inclination-type resonance, but eventually resonance overlap occurs. Tethys and Dione both exit the resonance with substantial eccentricity and inclination.} 
\label{td4c}
\end{figure}

\begin{figure}[p]
\epsscale{0.7}
\plotone{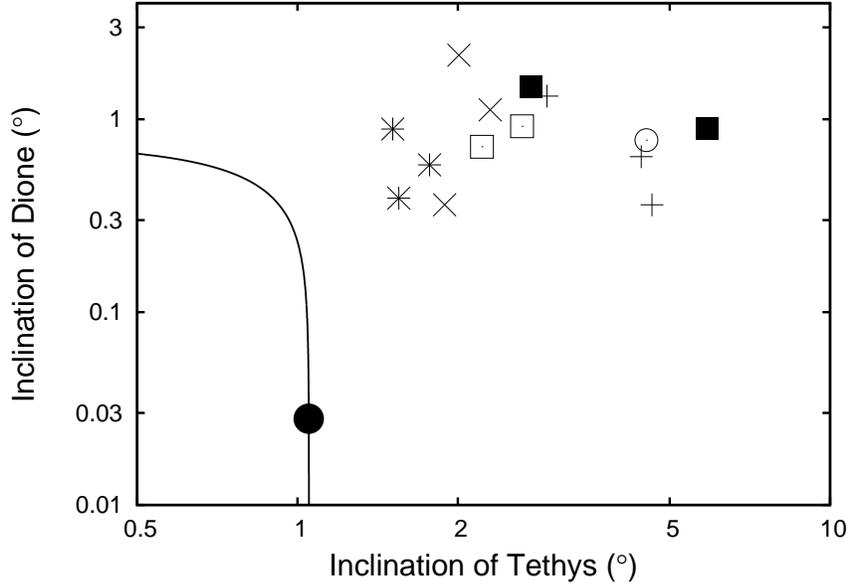}  
\caption{Outcome of several simulations of the Tethys-Dione 3:2 resonance crossing. The plot shows the post-resonance inclinations of Tethys (x-axis) and Dione (y-axis). Crosses stand for simulations with initially circular and equatorial orbits, while x's and asterisks plot simulations with circular orbits and initial $i=0.3^{\circ}$ and $i=1^{\circ}$ for Tethys, respectively. Coplanar simulations with initial $e=0.01$ for Tethys and Dione are plotted by solid and open squares, respectively. A simulation with coplanar and circular orbits but with $Q/k_2=200$ for the moons is plotted with an open circle (most other simulations assumed $Q/k_2=2000$ for both satellites, and we used $Q/k_2=4000$ for Saturn throughout). Multiple simulations with the same initial eccentricities and inclinations had slightly different starting semimajor axes for Tethys. The observed free inclinations of Tethys and Dione are marked with a solid circle, and states reachable by redistributing out-of-plane angular momentum between the two moons are plotted by a solid line. Neither the present orbits not their combined angular momentum deficit are matched by any of the simulations plotted here.} 
\label{scatter}
\end{figure}

The period ratio of Dione and Tethys is currently about 1.45, so, as Tethys evolves outward much faster than Dione, these two moons should have crossed their mutual 3:2 resonance at some point in the past. In this section we use the symplectic numerical integrator {\sc simpl} to explore the Tethys-Dione 3:2 resonance crossing. The benefit of the numerical approach is that all sub-resonances, including the inclination-related ones, are fully accounted for. Figure \ref{td4c} shows a typical evolution of eccentricities and inclinations of Tethys and Dione during the resonance crossing, assuming that both moons' orbits were close to circular and equatorial before the resonant encounter. The moons are originally captured in a sub-resonance which leads to a monotonic growth of the inclination of Tethys. Once Tethys's inclination grows to about 4$^{\circ}$, the resonance becomes chaotic and eccentricities and inclinations of both moons vary rapidly. Eventually the resonant lock breaks, leaving both orbits significantly eccentric and inclined. While the eccentricities can damp due to eccentricity tides, the inclinations of Saturn's mid-sized moons are not expected to evolve appreciably over the age of the Solar System \citep{che14}. The final inclination of Dione is always substantial, typically comparable to a degree (Fig. \ref{scatter}), while the observed inclination of Dione is only 0.028$^{\circ}$ (Table \ref{elem}).

\begin{figure}
\epsscale{0.8}
\plotone{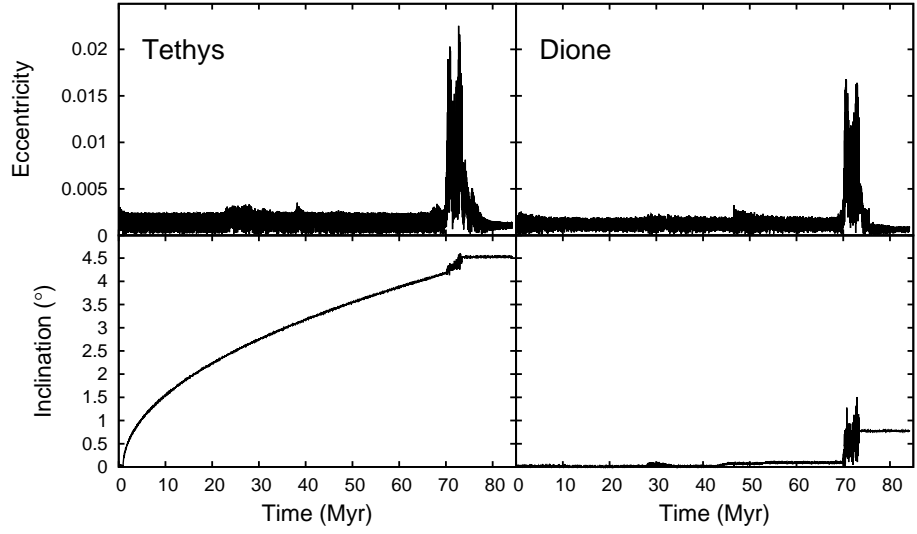}  
\caption{Same as Fig 1, only with $Q/k_2=200$ assumed for both moons.} 
\label{td8b3}
\end{figure}

\begin{figure}
\epsscale{0.8}
\plotone{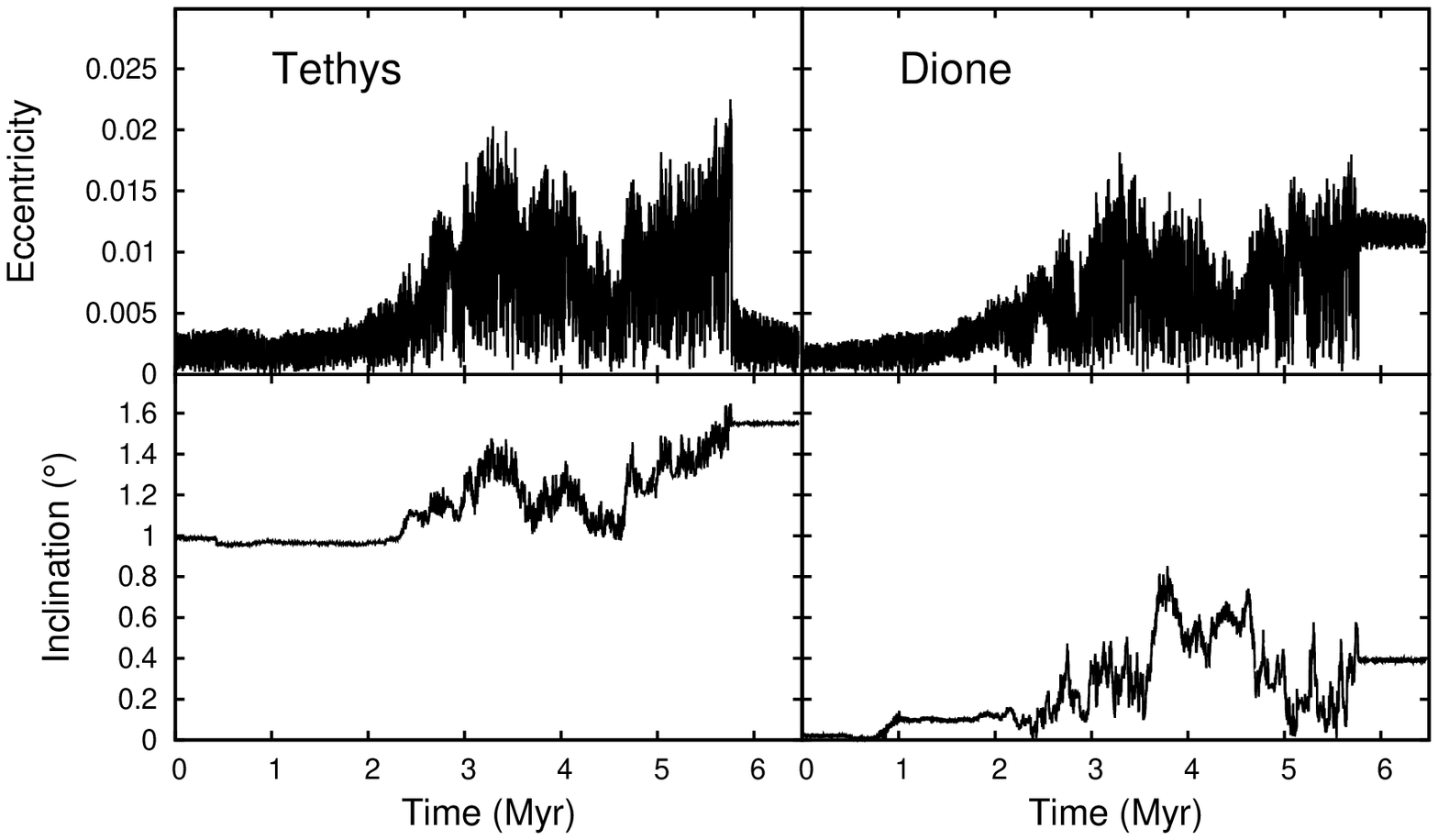}  
\caption{Same as Fig. 1, only with initial $i=1^{\circ}$ assumed for Tethys.} 
\label{td5}
\end{figure}

\begin{figure}[h]
\epsscale{0.8}
\plotone{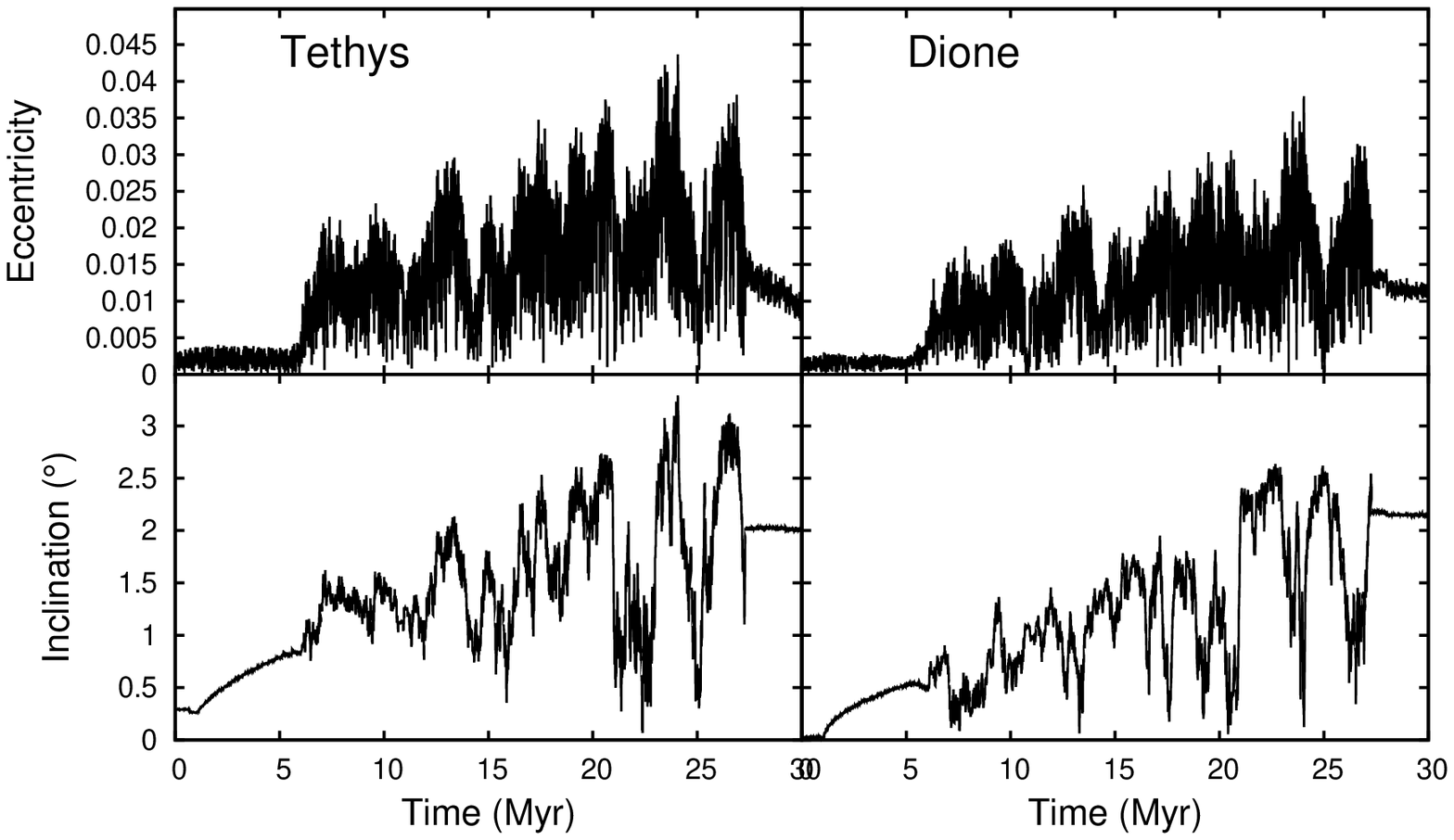}  
\caption{Same as Fig. 1, only with initial $i=0.3^{\circ}$ assumed for Tethys.} 
\label{td7a}
\end{figure}

We used $Q/k_2=4000$ for Saturn in almost all the simulations is this paper, which is equivalent to $Q=1360$, as Saturn's Love number is estimated to be $k_2$=0.341 \citep{gav77}\footnote{\citet{lai15} find $k_2=0.390 \pm 0.024$ for Saturn; this Love number would require $Q$ of Saturn in our simulations to also be 10-20\% larger}. For the satellites, we used $Q=100$ and $k_2=0.05$ in the simulation shown in Figure \ref{td4c}, which implies at least some melting, as expected when eccentricities reach a few percent \citep{che08}. A more detailed model would have tidal parameters varying with tidal heating, but our integrator did not include that capability. Instead, we explored a range of constant tidal parameters for the satellites. Given that Enceladus seems to currently have very low $Q/k_2$ \citep[if Enceladus's orbit is in equilibrium as proposed by ][]{lai12}, we also integrated the resonance crossing using $Q=100, k_2=0.5$ (Figure \ref{td8b3}). Despite strong tidal dissipation, the inclination of Dione is still excited during the crossing.  One can in principle imagine an even stronger tidal response from Tethys or Dione (with $Q/k_2 \simeq 10$), but these would cause Tethys to move inwards due to satellite tides being faster than the outward migration due to Saturn tides (for eccentricities of the order $e=0.01$). Such a reversal of migration would effectively prohibit Tethys from crossing the resonance with Dione and reaching the present configuration (cf. Eq. \ref{tides}). 

We found it impractical to directly model resonance crossing in the ``classical" case when Q=18,000 for Saturn, as the integrations would need to be extended to many hundreds of Myr. However, chaotic excitation of both the eccentricity and inclination is expected to be even stronger when the resonance is crossed more slowly. If we assume a larger tidal Q for Saturn, a wider range of tidal responses from Tethys would overwhelm Saturn's tides and lead to divergent migration, making the resonance an impassable barrier for Tethys. The ratio between the migration rate due to tidal dissipation within the satellite and the migration rate due to dissipation within the planet is: 
\begin{equation}
{{\dot a}_{sat} \over {\dot a}_{pl}} = - 7 e^2 {Q^{*}_{pl} \over Q^{*}_{sat}} \Bigl({ M_{pl} \over M_{sat} }\Bigr)^2 \Bigl({ R_{sat} \over R_{pl} }\Bigr)^2
\label{tides}
\end{equation}
where subscripts ``sat" and ``pl" refer to the satellite and the planet, $M$, $R$ and $e$ are respectively mass, radius and orbital eccentricity, while $Q^{*}=Q/k_2$ is the measure of tidal dissipation. For $e=0.01$ (typical of eccentricities excited by the Tethys-Dione resonance) and the size and mass of Tethys, we get $Q^{*}_{pl}/Q^{*}_{sat}$=0.03  for outward migration to stop. If $Q^{*}_{pl}=4000$, then $Q^{*}_{sat} > 120$ is necessary for Tethys to continue outward migration. This limit is comparable to the current tidal response of Enceladus inferred by \citet{lai12} by assuming an equilibrium. However, if $Q^{*}_{pl}=50,000$, then $Q^{*}_{sat} > 1500$ for the migration to stay convergent, which likely excludes any major melting within Tethys, despite significant eccentricity (if $Q=100$ then $k_2$ would have to be less than $0.07$, indicating a relatively rigid body). Without going into modeling the actual physical response of the satellite, we can just use the above considerations to exclude the parameter space with $Q^{*}_{pl}/Q^{*}_{sat} > 0.03$ from our consideration, as those parameters would forever trap Tethys on the inner side of the 3:2 resonance with Dione. The only exception is $Q^{*}_{pl} > 2 \times 10^5$, in which case the resonance was never crossed, regardless of satellite tides. However, this extremely slow evolution rate is about 40 times too slow to account for the heating of Enceladus, and we prefer the interpretation that the resonance was never crossed because the moons are young.

It is possible that the inclination of Tethys was already excited before it encountered the resonance with Dione. Analytical considerations open a possibility that a pre-inclined Tethys could have conceivably avoided capture into the inclination-type resonance evident in Fig. \ref{td4c} and the associated inclination increase. We tested this possibility using initial inclinations of 1$^{\circ}$ and 0.3$^{\circ}$ for Tethys in multiple simulations. Figs. \ref{td5} and \ref{td7a} each plot the results of one simulation with initial $i=1^{\circ}$ and $i=0.3^{\circ}$ for Tethys, respectively. For initial $i=0.3^{\circ}$ resonance capture still happens, although it usually lasts a much shorter time than in Fig. \ref{td4c}, while for initial $i=1^{\circ}$ the moons enter the chaotic behavior at the very beginning of the resonance crossing. In all cases the resulting final inclination of Dione is much higher than the observed one (Fig. \ref{scatter}). We also conducted simulations that start with one moon's eccentricity being $0.01$. While simulations with initially eccentric Dione did lead to the onset of chaos earlier in the simulation, those with initially eccentric Tethys show no difference from the circular case, chiefly because most of this initial inclination damps by the time chaos is triggered (despite $Q/k_2=10^4$ being used in this run). In all cases the inclinations of both Tethys and Dione are excited well in excess of the observed values (Fig. \ref{scatter}).

The numerical experiments shown in this section suggest that the inclinations of both Tethys and Dione are excessively excited by the crossing of their 3:2 resonance, regardless of the assumed initial conditions and tidal parameters. In the next section we will discuss possible resonances that could have affected Dione and Tethys later on, but they cannot make a past 3:2 resonance crossing compatible with the present orbits. The solid line in Fig. \ref{scatter} plots the combinations of inclinations that have the observed out-of-plane angular momentum deficit\footnote{Angular momentum deficit is defined as $\Sigma m_i \sqrt{a_i} (1 - \sqrt{1-e_i^2}) (1-\cos{i_i})$, and is approximately $\Sigma {1 \over 2} m_i \sqrt{a_i} (e_i^2+\sin^2{i_i})$ for small $e$ and $i$.} (AMD), and they are all well short of the inclination AMD seen among the simulation outcomes. Capture into additional mean-motion resonances could only increase the AMD of Tethys and Dione, while the secular resonance involving these two moons (see next section) would conserve AMD associated with inclinations. Likewise, chaotic interaction within a mean motion resonance is extremely unlikely to decrease the combined inclinations, but tends to redistribute inclinations on short timescales and increase AMD over time. This applies to subsequent resonances with Enceladus and Mimas; Mimas does have a large inclination but small AMD compared with Tethys due to its small mass, and could not have absorbed out-of-plane AMD from Tethys and Dione. The fact that both Tethys and Dione are always excessively inclined after their 3:2 MMR crossing is a strong constraint on their relative locations at formation, and we will discuss the implications in the following sections.

\section{Dione-Rhea 5:3 Resonance and Tethys-Dione Secular Resonance}

In the relative chronology of resonance crossings in the Saturnian system, the Dione-Rhea 5:3 mean-motion resonance would happen shortly after the 3:2 Tethys-Dione resonance. We have concluded in the previous section that the passage through 3:2 Tethys-Dione resonance did not happen, and here we explore the possibility that Tethys formed outside the 3:2 resonance with Dione and Dione formed interior to the 5:3 resonance with Rhea, and that Dione and Rhea have since crossed their 5:3 resonance.

\begin{figure}[!ht]
\epsscale{1.1}
\plotone{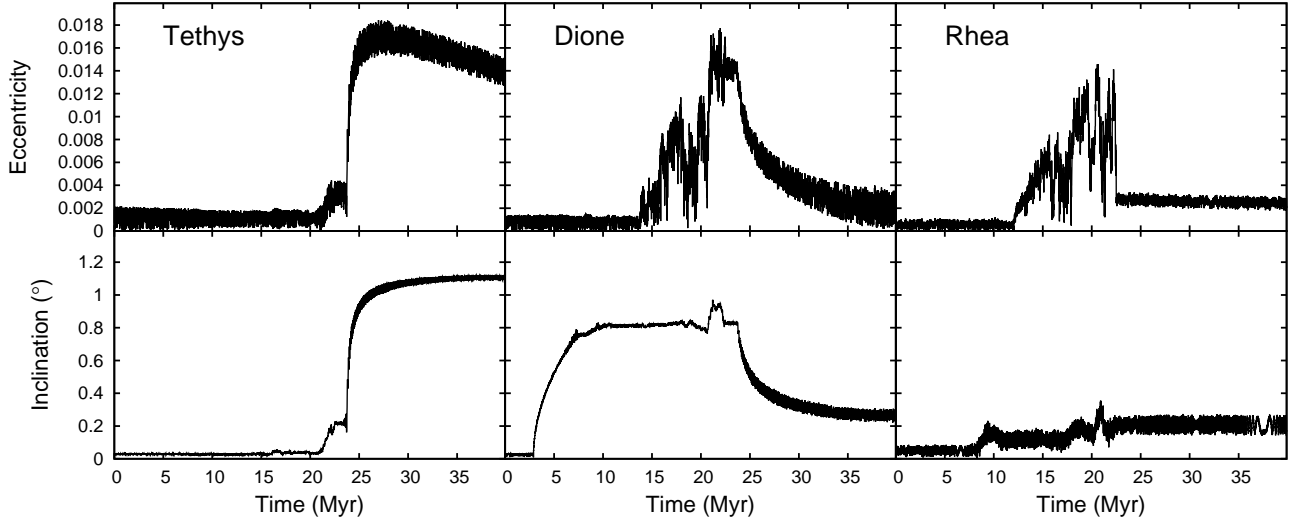}  
\caption{A simulation of Dione and Rhea encountering their mutual 5:3 resonance, followed by the Tethys-Dione secular resonance. Tidal response of Saturn was $Q/k_2=4000$, while the satellites were assumed to have tidal parameters $Q=100$ and $k_2=0.05$. Tethys, Dione and Rhea are assumed to have low eccentricities and inclinations before the resonance crossing. At first Dione is captured in the $i_1^2$ 5:3 inclination-type resonance, but eventually resonance overlap occurs. Dione and Rhea both exit the resonance with substantial eccentricity and inclination. At about 24 Myr, Tethys and Dione enter the secular resonance, transferring eccentricity and inclination from Dione to Tethys.} 
\label{dr3}
\end{figure}

Figure \ref{dr3} shows one simulation of the 5:3 Dione-Rhea MMR crossing. In the beginning Dione is captured in the inclination-type resonance, which eventually breaks and a chaotic phase follows, during which the eccentricities and inclinations of both bodies are excited. This part of the evolution is expected, and while it could explain the excitation of Rhea's inclination ($i=0.35^{\circ}$), it is seemingly at odds with the very low inclination of Dione ($i=0.03^{\circ}$). While the eccentricities are also excited, they can be expected to damp later through satellite tides. However, at 24 Myr we see a striking exchange of eccentricities and inclinations between Tethys and Dione. Both the eccentricity and inclination of Tethys get excited this way, while those of Dione are reduced. We find that this unexpected dynamical event is caused by a previously unidentified Tethys-Dione secular resonance.

The secular resonance seen in these simulations has the resonant argument $\Psi = \varpi_2-\varpi_1 +\Omega_2-\Omega_1$, where $\varpi$ and $\Omega$ are the longitudes of the pericenter and the node, and the subscripts 1 and 2 refer to Tethys and Dione, respectively. The pattern of signs in the resonant argument indicates that in this resonance the eccentricity and inclination of each moon should be correlated, and anti-correlated with those of the other moon, consistent with numerical integrations showing Tethys getting more inclined and eccentric, with Dione's orbit becoming more circular and planar. The resonance appears to break when either the eccentricity or inclination of Dione reaches zero, with the latter case being able to explaining the very low present inclination of Dione. Note that in Fig. \ref{dr3} it is eccentricity, rather than inclination, that reaches zero first and breaks the secular resonance.

The type of secular resonance seen here affecting Tethys and Dione has not, to our knowledge, been described in the literature. Therefore, here we briefly discuss why it arises soon after the Dione-Rhea 5:3 mean motion resonance. The resonant argument $\Psi$ is a combination of the arguments $\varpi+\Omega$ for each moon, which are typically very slowly varying for a regular satellite, as ${\dot \varpi} \simeq - {\dot {\Omega}}$ for low-$e$, low-$i$ orbits chiefly perturbed by planetary oblateness. Usually, the leading term in $\dot{\varpi}+\dot {\Omega}$ is $9/4 J_2^2 (a/R)^4 n $ \citep{md99}, where $J_2$ is the standard oblateness moment and $R$ the radius of Saturn, and $a$ and $n$ are the moon's semimajor axis and mean motion. This result does not depend on the definitions of orbital elements used (geometric vs. osculating), which affect individual expressions for ${\dot {\Omega}}$ and ${\dot {\varpi}}$ but cancel out in their sum \citep{gre81}. This term currently leads to periods of about 5000~yr and 20,000~yr for $\varpi+\Omega$ of Tethys and Dione, making $\dot{\Psi} < 0$. Clearly, another source of precession is necessary to bring these two moons into a secular resonance.

Precession due to oblateness is also affected by eccentricity and inclination. From \citet{dan92}, we have:
\begin{equation}
{\dot \varpi} + {\dot \Omega} = {\dot \omega}+{2 \dot \Omega} = {3 n J_2 \over a^2 (1-e^2)^2} \Bigl( 1 - {5 \over 4} \sin^2{i} - \cos{i} \Bigr) \simeq { 3 n J_2 \over a^2} \Bigl( - {3 \over 4} \sin^2{i} + O (i^4)\Bigr)   
\end{equation}
Where $\omega = \varpi - \Omega$ is the argument of pericenter, and $e$ and $i$ are the moon's orbital eccentricity and inclination (to the planet's equator). For $i=1^{\circ}$, we find that the quantity in parenthesis dependent on inclination is only $-2 \times 10^{-4}$, which for Tethys would produce retrograde precession of ${\dot {\Omega}}+{\dot {\varpi}}$ with a period of 11,000~yr. This is a potentially significant term, but it would add a negative term to ${\dot {\Psi}}$ before the resonance is encountered, as it is Dione, rather than Tethys, that must be inclined before the resonance capture occurs. Therefore this term cannot be the cause of the resonance being encountered in the first place. However, inclination-dependent terms may be important for maintaining the resonance, as they will slow down the excess apsidal precession of Tethys as it acquires inclination. 

Since the secular resonance is encountered near Dione's 5:3 resonance with Rhea, it appears that near-resonant terms are important for establishing the secular commensurability. With this being the second-order resonance, six different resonant terms can affect the precession of Dione. After some numerical testing, we have established that this additional precession is sensitively dependent on Dione's distance from the outermost sub-resonance (associated with the $e_2^2$ term) and is also strongly dependent on Dione's eccentricity. These terms appear to accelerate the precession of Dione's longitude of pericenter, but do not affect the node as much, producing a significant change in the secular resonant argument $\Psi$. 

%We also found that analytical estimates of near-resonant precession term are both difficult and inaccurate, and we will use only direct numerical methods in the present paper.

\begin{figure}
\epsscale{1.}
\plotone{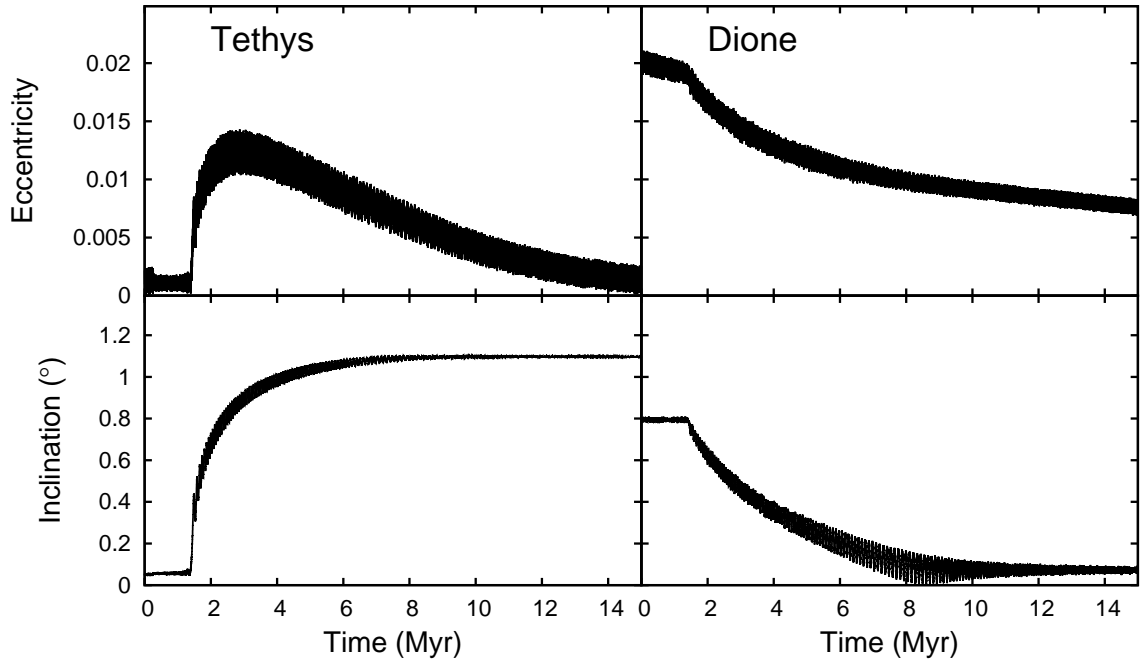}  
\caption{A simulation of Tethys and Dione encountering the secular resonance. In this simulation the moons were initially placed immediately outside the Dione-Rhea 5:3 mean-motion resonance, with eccentricities and inclinations close to those required for reaching their observed inclinations through the secular resonance.} 
\label{secres8}
\end{figure}

Figure \ref{secres8} shows a simulation where Dione and Rhea are placed just outside their 5:3 resonance (i.e. after an assumed 5:3 resonance passage), with Tethys having a near-circular and planar orbit and Dione having $e=0.02$ and $i=0.8^{\circ}$ (Rhea, not plotted, had initial $e=0.006$ and $i=0.3^{\circ}$). About 1.4 Myr into the simulation the secular resonance is established, and it lasts until about 8.5 Myr, when the average inclination of Dione drops below $0.1^{\circ}$. While this is a somewhat artificial set-up (we assume the outcome of the 5:3 Dione-Rhea resonance, without simulating it), it clearly shows the potential of the secular resonance to produce the current orbits of Tethys and Dione, with the right initial conditions\footnote{Secular resonance capture does not require a small initial inclination of Tethys (its inclination only has to be lower than the current value so it can absorb Dione's inclination), but we will assume that the initial tilt was close to zero for sake of simplicity. We have already concluded that the Tethys-Dione 3:2 resonance did not happen, and we are not aware of any other possible sources of Tethys's inclination.}. We find that resonance capture is almost certain (assuming $Q/k_2=4000$) if the eccentricity of Dione is larger than $0.01$. The current low inclination of Dione indicates that, if it was reduced by this secular resonance, the resonance had to break when the inclination of Dione reached zero (as opposed to the situation shown in Fig. \ref{dr3}, where eccentricity reaches zero first). As $\dot{e}_2=\dot{i}_2$ in this secular resonance, and the angular momentum is conserved, this means that the initial $e_2 > (a_1/a_2)^{1/4} (m_1/m_2)^{1/2} \sin{i_1} = 0.012$. This gives us a requirement for post-5:3 resonance orbits of Dione and Rhea: $e_2 > 0.012$, $e_3$ unconstrained, $i_2=0.7^{\circ}$ and $i_3=0.33^{\circ}$ (current value). We assume low eccentricity and inclination for Tethys at this time, as higher eccentricities of Tethys often lead to greater chaos and sometimes overlap between the Tethys-Dione 3:2 and Dione-Rhea 5:3 resonances, which is long-lasting, hard to break and excites inclinations beyond the observed values. 

\begin{figure}
\epsscale{1.}
\plotone{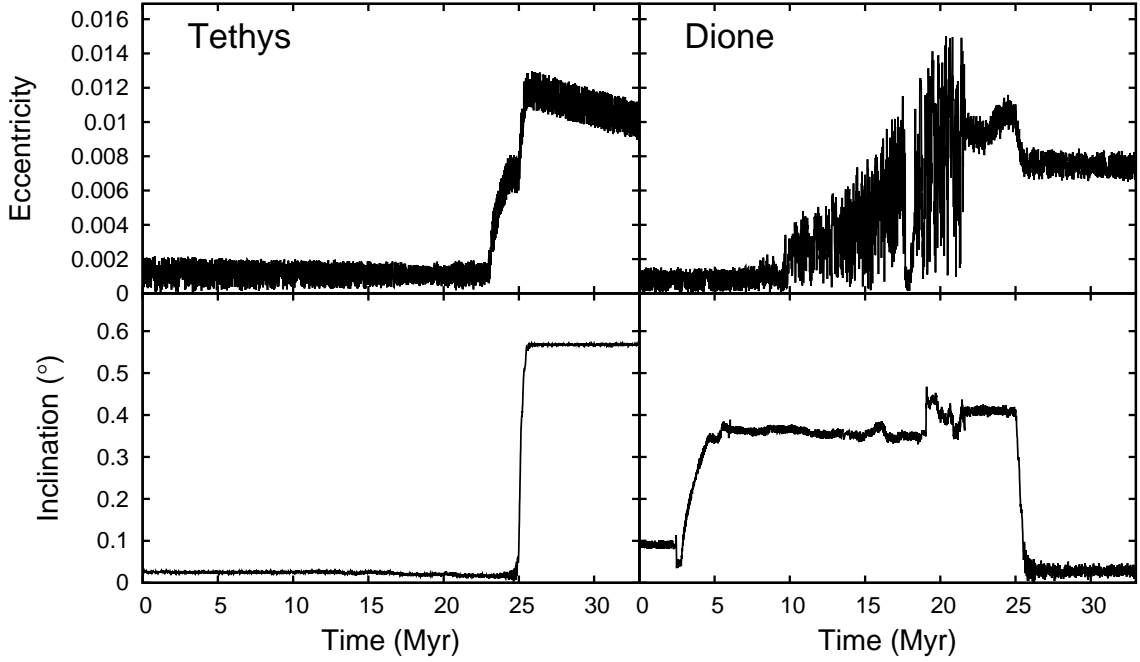}  
\caption{A simulation of Tethys and Dione encountering the secular resonance immediately outside the Dione-Rhea 5:3 mean-motion resonance. This is one of the set \#5 simulations plotted in Fig. \ref{scatter2} ($Q/k_2=10^4$ for the moons and 4000 for Saturn, and Titan's eccentricity $e_T=0.04$). Due to its non-zero initial inclination Dione jumps through the $i_2^2$ resonance but is then captured in the next sub-resonance, which breaks at $i_2=0.04$. in this simulation we see temporary capture into a three-body sub-resonance of the Dione-Rhea 5:3 MMR that also involves Tethys, with the argument $5 \lambda_3 -3 \lambda_2 - \varpi_2 -\varpi_1$. By increasing the eccentricities of Tethys and Dione, such subresonance captures typically make the secular resonance capture more likely.} 
\label{dr17a3}
\end{figure}

While the required initial conditions for the secular resonance fall within the range of stochastic outcomes of the 5:3 Dione-Rhea resonance, we could not find a combination of initial conditions and tidal parameters that lead to such a state with high probability. We find that the initial capture into the $i_2^2$ subresonance of the 5:3 MMR (involving only the inclination of Dione) is certain for very low initial inclinations ($i_2 \simeq 0.02^{\circ}$). This sub-resonance, however, tends to break only at $i_2=1.4^{\circ}$, which is well above the value required for the secular resonance. An initial inclination $i_2 =0.1^{\circ}$ leads to a jump through the the $i_2^2$ sub-resonance and capture into the the $i_2 i_3$ subresonance, in which both the inclinations of Dione and Rhea grow. This resonance usually breaks in the $i_2=0.2-0.7^{\circ}$ range, approaching the target inclination needed for the secular resonance, while the inclination of Rhea reaches $i_3=0.1-0.4^{\circ}$, approaching the current values. Figure \ref{dr17a3} shows one simulation using initial $i_2=0.1^{\circ}$ that results in a very low inclination of Dione and a substantial inclination of Tethys after the secular resonance.

\begin{figure}
\epsscale{1.}
\plotone{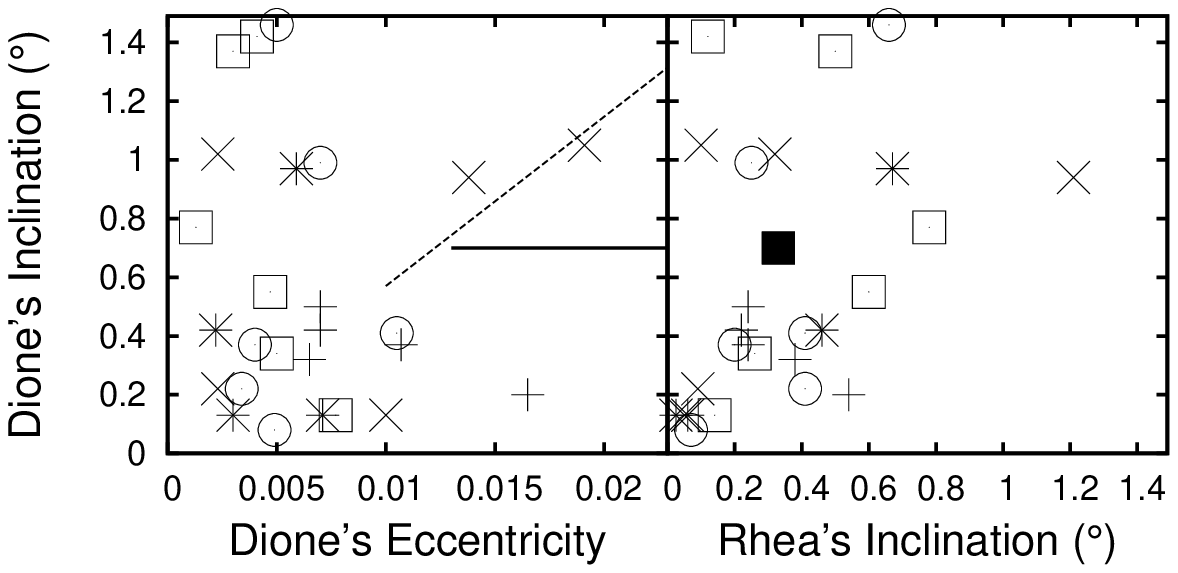}  
\caption{Outcome of a number of numerical experiments of the Dione-Rhea 5:3 mean motion resonance crossing, also including Tethys and Titan. The values plotted here are representative of the time just after the mean-motion resonance and before the Tethys-Dione secular resonance (where applicable). We ran five sets of eight simulations using initial $i_2=0.1$ for Dione and different tidal parameters, and final eccentricities and inclinations were plotted here for cases where the mean-motion resonance was crossed in less than 50 Myr (or 70 Myr for Set \#2). Sets \#1 (crosses), \#2 (Xs) and \#5 (circles) had $Q/k_2=10^4$ for the moons, while sets \#3 (asterisks) and \#4 (squares) had $Q/k_2=200$ for all three icy satellites. Set \#2 (Xs) had $Q/k_2=8000$ for Saturn, while all other sets had $Q/k_2=4000$ for the planet. The first three sets had $e_4=0.007$ for Titan, while sets \#4 and \#5 had more realistic initial $e_T=0.04$. Secular resonance typically follows if $e_2>0.01$, and produces very low inclinations for Dione if the initial $e_2 > \sin{i_2}$ (i.e. for initial conditions below the dashed line). The thick black line and the solid square plot the initial conditions for the secular resonance that are consistent with the moons' present inclinations.} 
\label{scatter2}
\end{figure}

Figure \ref{scatter2} shows the outcome of several sets of simulations starting with $i_2=0.1^{\circ}$, with all other eccentricities and inclinations (except those of Titan) being close to zero. It is clear from the figure that the amount of excitation needed to explain the present system roughly matches the outcomes of the Dione-Rhea MMR. While no single simulation exactly reproduces the current orbits, very low inclinations of Dione and significant inclinations of Tethys are often produced through the secular resonance.

While the mechanism involving the 5:3 Dione-Rhea resonance followed by the Tethys-Dione secular resonance is a very promising avenue for producing the observed inclinations, we were not able to achieve this outcome with any reasonable probability using {\sc simpl}. One possible explanation is that the process is so stochastic that we cannot expect to exactly reproduce the real history of the system. We think it is still too early to reach such a conclusion, and hope to explore this question in future work. The major limitations of our present approach include: 1) we assume that the tidal properties of the satellites are constant over time, despite varying eccentricities and tidal heating; 2) we assume Saturn's tidal $Q$ is frequency-independent, which has implications for the exact timing of recent resonance crossings; 3) while our integrator allows it, we did not explore different initial tidal parameters for the three moons (i.e. we assumed they were all equally rigid or dissipative, which is likely to be wrong). A clear step forward would be incorporation of a geophysical model into the integrator that would adjust tidal parameters depending on the tidal heating and other processes the moon is subject to.

The chief conclusion of this section is that the past 5:3 mean-motion resonance between Dione and Rhea likely did happen, unlike the 3:2 Tethys-Dione resonance discussed in the previous section. This gives us a very well-defined relative age of Tethys, Dione and Rhea, which is related to their absolute age through the controversial tidal $Q$ of Saturn. As for the absolute ages, the classical $Q=18,000$ \citep{mey07} and the ``fast" $Q=1700$ \citep{lai12} would place the formation of Tethys, Dione and Rhea at about 1.1 Gyr and 100 Myr ago, respectively, with both dates being much younger than the age of the Solar System. In the following sections, we will briefly explore the possible dynamical mechanisms that could have led to the late formation of the present set of mid-sized icy moons of Saturn. 

%There are two possible interpretations of this constraint. Either Saturn has Q$>80,000$, and the moons' orbits evolved only by a small amount over the age of the system, or the moons are significantly younger than 4.5 Gyr. Given strong (presumably tidal) heating of Enceladus (40 times above the equilibrium if Q=80,000), we favor the hypothesis that the moons are young. This experiment does not give us the absolute age of the moons, but based on astronomer reported by \citet{lai12} and the state of the Titan-Hyperion resonance \citep{cuk13}, we think that the most likely age of the inner moons is about 100 Myr or less.

\section{Evolution of the Protosatellite Disk}

As we are proposing that the moons of Saturn (out to Rhea) formed much later than the planet, we need to address the plausibility and basic physics of such an event. First of all, the flatness of Saturn's regular satellite system indicates formation from a thin disk. Since we think that Tethys, Dione and Rhea all formed close to their present orbits, this disk would need to extend out to about 10 Saturn radii and comprise the combined mass of the mid-sized moons. Since the Titan-Hyperion resonance appears very old \citep{gre73, cuk13}, we assume that Titan (and other more distant moons) predated this protosatelite disk that Tethys, Dione and Rhea formed from. This does not necessarily mean Titan must be primordial, as it may have originated during another, more ancient satellite instability event \citep{ham13}, which is outside the focus of this paper. As Saturn has youthful and dynamic rings surrounded by numerous small moons \citep{cuz10}, we propose that the rings also formed in the same event as Tethys, Dione and Rhea. This means that the protosatellite disk discussed here extended all the way inward to the Roche limit (possibly due to viscous spreading). Unlike a primordial circumplanetary disk, this disk would be comprised exclusively of solids, save for short-lived vapor created in collisions. In this Section we discuss the likely lifetime and fate of this disk, while in Sections 7 and 8 we will address the origin of the disk.

\citet{bur98} used numerical simulations to study the fate of ejecta in the Saturnian system in order to constrain possible hazards for the Cassini spacecraft. They found that the moons accrete any material initially found on crossing orbits in mere centuries \citep[also see][]{alv05}, and theoretical predictions of their growth rate from debris are also on the order of millennia \citep[cf.][]{gol04, chi13}. We can use the expression in \citet{chi13} \citep[also see][]{gol04} to calculate the coagulation time:
\begin{equation}
t \simeq {1 \over F_{grav}} {\rho_{core} R_{core} \over \sigma_{solid} \Omega}
\end{equation}
where $\rho$ and $\sigma$ are the volume and surface densities of a solid core and the disk, respectively, and $F_{grav}$ is the gravitational focusing factor (we will assume $F_{grav}=1$). If we estimate $\sigma_{solid}=4 \times 10^{21}{\rm \ kg }\ \pi^{-1} (6 \times 10^8 {\rm \ m})^{-2}  \simeq 4000 {\rm \ kg \ m^{-2}}$, $\rho_{core}=1200 {\rm \ kg \ m^{-3}}$ and $\Omega = 2 \times 10^{-5} {\rm s^{-1}}$  (corresponding to a 4-day orbital period), we get 
\begin{equation}
t \simeq 0.5 {\rm \ yr \ km^{-1}} R_{core} 
\end{equation}
Therefore, it would take less than 1000 years to grow 1000 km satellites, in agreement with the numerical integrations of debris removal by \citet{bur98}. However there are serious problems with this approach. The first is that the present satellites have well-separated Hill spheres and would therefore not sweep up all of the material between them. Assuming a uniform disk, we can estimate the number ($n$) and size ($r$) of the satellites assuming that they are separated by 5 Hill radii: 
\begin{equation}
r= 5 n R_{hill} =  5 n r \Bigl({M \over 3 M_{pl}}\Bigr)^{1/3} =  5 n r  \Bigl({M_{disk} \over 3 n M_{pl}}\Bigr)^{1/3}
\end{equation}
so therefore
\begin{equation}
n= 5^{-3/2} \Bigl({ 3 M_{pl} \over M_{disk}} \Bigr)^{1/2} \simeq 60,
\end{equation}   
where we have assumed $M_{disk} \simeq M_{Rhea} + M_{Dione} + M_{Tethys}$. This makes the mass of a typical protosatellite $M=M_{disk}/n=0.7 \times 10^{20}$~kg, comparable to that of Enceladus. These protosatellites would take significantly longer to merge into the larger moons we observe today. The mechanism for their destabilization and mergings would likely be chaotic interaction between the moons, many of which would likely be in overlapping mean-motion resonances with several others. Additionally, there would be some fragmentation during their accretion and mergings, and rings of debris would likely occupy quasi-stable areas between neighboring moons' cleared zones. Interaction with this debris would also affect the protosatellites' orbits, leading to orbit crossings and mergers. Direct numerical modeling is clearly needed to explore the evolution of such a system. 

Another caveat with accretion is outside perturbations, mainly from Titan. As \citet{bur98} show, the orbits close to Titan's 3:1 and 2:1 resonances are significantly more perturbed than those in between mid-sized moons. This excitation would inhibit accretion, at least in some locations, potentially leading to a relatively long-lived debris ring outside the orbit of Rhea. Eventually this disk would spread and disperse due to inter-particle scatterings and collisions. \citet{cri12} find that the timescale for the lifetime of planetary rings, based on numerical experiments, can be expressed as:
\begin{equation}
T_{disk} = \Bigl({\pi \over 52 (R_H / R_{part})^5}\Bigr) {M^2_{pl}\over \sigma^2_{disk} r_{disk}^4} T_{orbit}
\end{equation}
where $R_{part}$ and $R_H$ are a particle's physical and Hill radius, respectively. Assuming the disk radius of $r=6 \times 10^8$~m (ten Saturn radii) and a density of 500~kg~m$^{-3}$ for disk particles, $(R_H/R_{part}) \simeq 6$ and $T_{orbit}$=5 days,
\begin{equation} 
T_{disk} = 3 \times 10^{11} {\rm yr} {1 \over \sigma^2_{disk} [{\rm \ kg \ m^{-2}}]^2}
\end{equation}
If we assume a surface density $\sigma_{disk}=4000 {\rm \ kg \ m^{-2}}$, the spreading time according to the above formula is about 2000 yr. After the surface density decreases by a factor of a few, the disk can last tens of kyr. At that point it would still contain a mass on the order of $M=10^{21}$~kg, making it possible to force protosatellites (or satellite fragments) of comparable mass to migrate as the disk is spreading. While the material on the inner edge would likely be accreted by Rhea, the large gap between the inner system and Titan would leave space for any protosatellites (or fragments of disrupted satellites) to migrate and potentially interact with Titan through mean-motion resonances. 

\section{Excitation of Titan's Eccentricity}

\begin{figure}
\epsscale{0.8}
\plotone{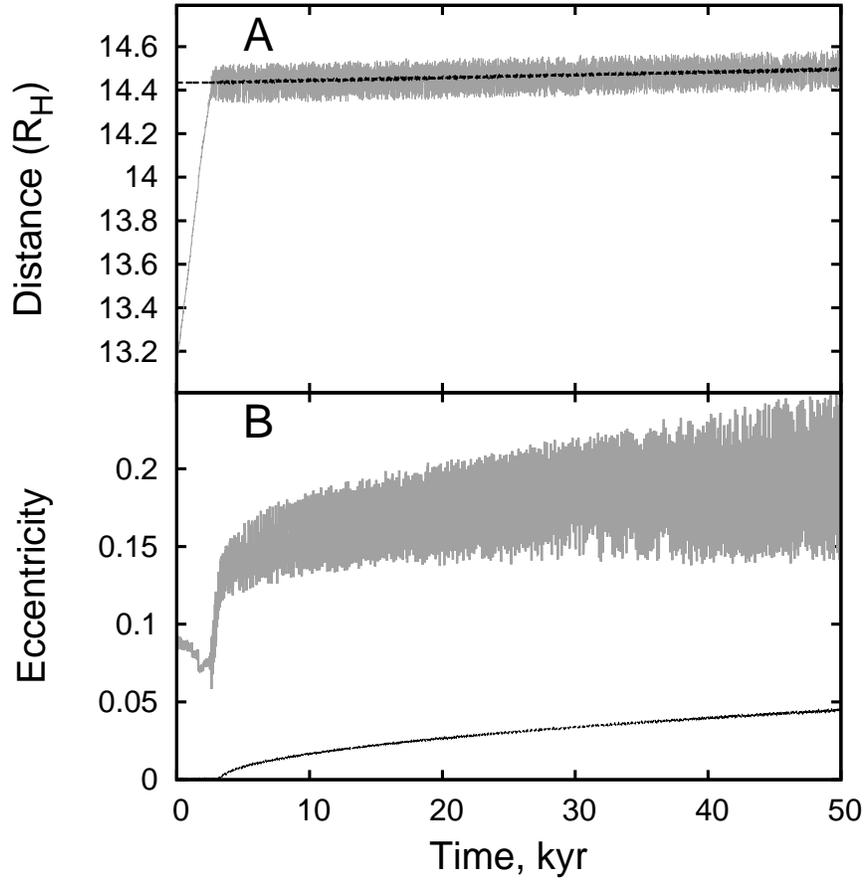}  
\caption{Excitation of Titan's eccentricity through a 3:5 resonance with a rapidly migrating moon of mass $m=4 \times 10^{20}$~kg. Gray lines plot the orbital elements for the inner moon, while the black lines plot the elements of Titan. In panel A, the gray line shows the semimajor axis of the inner moon, while the black line plots the location of the 5:3 resonance with Titan.} 
\label{mig6}
\end{figure}

The outer part of the disk, outside the 3:1 resonance with Titan (and particularly around the 2:1 resonance at 12.7 Saturn radii), would experience strong forced eccentricity due to Titan, and may endure as a debris disk for a much longer time \citep{bur98}. A satellitesimal (or a fragment) that is ejected outwards may be able to migrate by interacting with the spreading disk, and become trapped in a mean motion resonance with Titan. The migration of such an object due to disk spreading could be much faster than that due to tidal migration and could excite Titan's eccentricity to the observed value in a relatively short amount of time. In the simulation shown in Figure \ref{mig6}, we introduced an artificial fast migration of a $4 \times 10^{20}$~kg moon (0.3\% of Titan's mass) starting around 13 Saturn radii, just outside Titan's 2:1 resonance. Our hypothesis is that there would be a spreading ring of excited, non-accreting debris around the location of the resonance. The initial eccentricity of the moon was set to $e=0.1$, as we would expect this moon's (or fragment's) orbit to be excited. The moon in question migrates out until it is caught in the 5:3 resonance with Titan. In the first 1000 years after resonance capture, it is the eccentricity of the inner moon that grows, indicating a Lindblad resonance (involving the smaller body's eccentricity). However, very soon the system settles into a ``co-rotation"-type resonance (involving the larger body's eccentricity) and stays in it for the rest of the simulation. This behavior is unexpected and is likely related to the relatively rapid rate of evolution, which stabilizes formally overlapping sub-resonances.

\begin{figure}
\epsscale{0.9}
\plotone{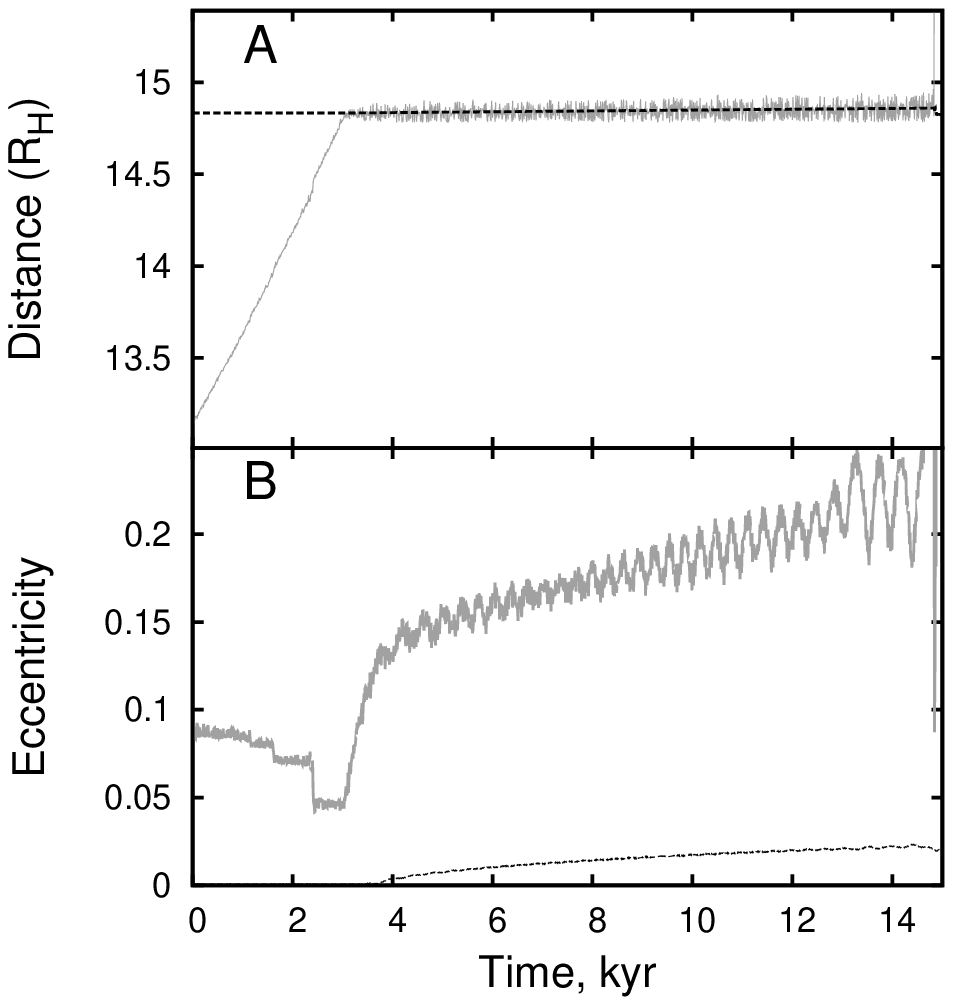}  
\caption{Excitation of Titan's eccentricity through a 5:8 resonance with a rapidly migrating moon of mass $m=8 \times 10^{20}$~kg. Gray lines plot the orbital elements for the inner moon, while the black lines plot the elements of Titan. In panel A, the gray line shows the semimajor axis of the inner moon, while the black line plots the location of the 8:5 resonance with Titan.} 
\label{mig5}
\end{figure}

\begin{figure}
\epsscale{0.9}
\plotone{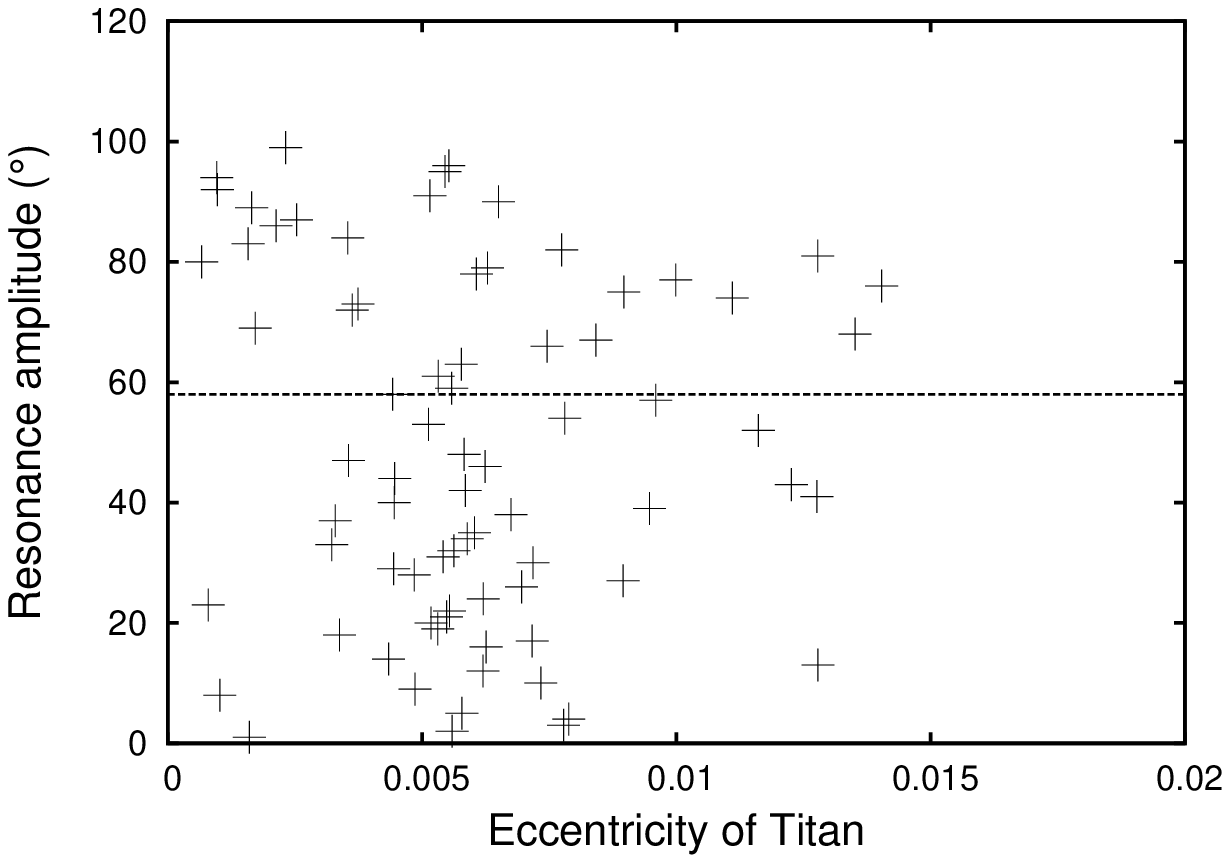}  
\caption{Maximum libration amplitudes of the Titan-Hyperion resonance argument $4 \lambda_H -3 \lambda_T - \varpi_H$ at the end of 76 10,000-year simulations of Titan and Hyperion following collisions between a scattered $9 \times 10^{20}$~kg satellite and Titan (see text). The maximum amplitude was calculated as the largest excursion from the libration center at $180^{\circ}$ over 10,000 years (the observed value is $59^{\circ}$; indicated by the dotted line). } 
\label{shotgun}
\end{figure}

In 50,000 years, the eccentricity of Titan grows to about 0.05, well above its observed value. Eventually, the resonance would break as the eccentricities of both bodies become high (the inner moon is indirectly affected by Titan's eccentricity), or the migration slows down and long-term chaos takes over. Figure \ref{mig5} shows a similar simulation involving a twice larger inner moon ($M=8 \times 10^{20}$~kg), and a similar capture into the 8:5 resonance with Titan. After the resonance breaks, the inner moon collides with either Titan or the disk.

Re-absorption of the protosatellite into the disk would have no consequences for Titan or Hyperion. However, close encounters followed by a collision with Titan would perturb the orbits of both Titan and Hyperion. If the Titan-Hyperion resonance is several Gyr old \citep{gre73, cuk13}, its continued existence might be in conflict with any destructive events in recent history. We therefore used {\sc complex} to run a set of 100 simulations of scatterings between Titan, Hyperion and a $9 \times 10^{20}$~kg satellite. The ``rogue" satellite was initially placed on an eccentric orbit with $a=13.2 R_S$ and $e=0.53$, with each run having a different longitude of pericenter or a different initial longitude for the inner satellite. Titan and Hyperion were initially placed deep in the resonance, with Titan on a circular orbit and Hyperion having $e=0.1$. In 85 out of one hundred 1000-year simulations using {\sc complex}, Titan and the inner moon collided, and in 76 of these Titan and Hyperion maintained a stable 4:3 resonance. The stability of the resonance was determined by integrating post-merger Titan and Hyperion for 10,000 years using {\sc simpl}.

The average orbits of Titan and Hyperion in the post-merger {\sc simpl} simulations are shown in Fig. \ref{shotgun}. The eccentricity of Titan is excited to about 0.01-0.02 on average (about half of its present value), requiring a different excitation mechanism (such as the mean-motion resonance described in this section) to explain its present orbit. The main (resonant) component of Hyperion's eccentricity stays close to 0.1, with the other (``slow") component being proportional to Titan's eccentricity. The resonant argument is excited, and the present libration semi-amplitude of $59^{\circ}$ is firmly within the ensemble of outcomes. 

We conclude that a resonant interaction with a moon as small as $4 \times 10^{20}$~kg, migrating due to disk spreading, can excite Titan's eccentricity to the current value. Even if this moon subsequently impacts Titan, the Titan-Hyperion mean-motion resonance is likely to survive this event. We numerically confirm resonance survival for a ``rogue moon" mass of $9 \times 10^{20}$~kg, and it is certain that collisions with smaller moons would have an even less disruptive effect on the Titan-Hyperion resonance. An eccentricity excitation only about 100 Myr ago would explain the surprisingly large free eccentricity of Titan despite its likely dissipative surface and interior \citep{sag82, soh95}.

\section{Resonances in an Ancient Satellite System}

In the previous sections we find that Saturn's present-day moons interior to Titan's orbit likely accreted much after the planet's formation, presumably from a disk, which covered the locations of the moons from Mimas to Rhea. As we discussed in Section 5, this disk would last only millennia, possibly somewhat longer in locations where Titan's perturbations inhibit accretion. Therefore we think that the disk was a temporary feature, and we propose that it originated in the destruction of a previous satellite system. While we argue that Saturn's present mid-sized satellite system is relatively young, its expected lifetime is measured in Gyr, \footnote{Using $Q=1700$ for Saturn, the next major resonance (Tethys-Dione 4:3 MMR) will not happen for at least 2 Gyr.} even in the case of ``fast tides" proposed by \citet{lai12}. Therefore, we argue that the most long-term arrangement of the mass currently in the mid-sized moons of Saturn is a system of satellites. As this previous system of satellites would have had the extent and total mass of the present mid-sized icy moon system, we will assume it also had most of its mass in the outermost two satellites, like Dione and Rhea today (this architecture is also most stable against tidal evolution).

Is there a way of destabilizing a mid-sized satellite system? We expect that any perturbations that can lead to large orbital eccentricities and moon-moon collisions would have to be resonant in nature. In this and the following sections, we will discuss several candidate resonances for destabilization of the putative previous-generation satellite system of Saturn. Assuming $Q \simeq 1700$ for Saturn, a Rhea-sized moon would evolve to about the present location of Rhea over 4.5 Gyr, while a somewhat smaller moon (like Tethys or Dione) would converge on positions very close to, but interior to,  Rhea's present orbit. This arrangement, with a Rhea-sized moon close to Rhea's present orbit, and a ``proto-Dione" on a converging interior orbit, is the starting point for our studies of past dynamical instability.

The simplest case would be a mean-motion resonance like that of Tethys and Dione described in Section 3. However, the relatively small masses of the inner moons make it impossible to excite their eccentricities to the high values required for orbit crossing (at least $e=0.1-0.2$). Tidal evolution while the moons are captured in a mean-motion resonance can in principle lead to monotonic eccentricity growth, but tidal dissipation within the moons can easily limit eccentricity growth for realistic physical parameters. A chaotic resonance between a large and a much smaller moon could excite the smaller moon's eccentricity enough to cause a collision, such a mismatched collision is unlikely to lead to disruption of a Dione or Rhea-sized moon \citep{lei12}.

Resonances between a mid-sized satellite and Titan have the potential of being more powerful, but are also unlikely to have been the cause of a past instability. We find that the 4:1 resonance with (non-eccentric) Titan at about 8 Saturn radii leads to temporary capture, after which the inner moon acquires an eccentricity of $e \leq 0.01$ and $i=0.3^{\circ}$. Another major resonance close to the mid-sized moons is the 3:1 with Titan, at 10 Saturn radii. This resonance is strongly chaotic, making the inner moon significantly eccentric and plausibly leading to instability. However, this resonance overlaps with the 4:1 resonance with Hyperion (which is in the 3:4 resonance with Titan). As Hyperion is small and does not damp eccentricity, a Rhea-sized moon in the 3:1 resonance with Titan quickly destabilizes Hyperion. Due to the survival of Hyperion we can exclude the 3:1 resonance with Titan as the source of a past instability.

%\section{Evection resonance}

Another set of resonances in the region inhabited by the mid-sized satellites is the so-called evection-type resonances. These are commensurabilities between a satellite's apsidal precession and the Sun's (or another distant perturber's) apparent planetocentric orbital period. Given Saturn's oblateness and heliocentric orbit, the strongest of these resonances happen for moons at about 8.3 Saturn radii, which have an apsidal precession period equal to Saturn's 29.5 year orbital period. Additional harmonics with evection happen at 6.9 Saturn radii (precession is 1/2 of Saturn's year) and 7.3 Saturn radii (precession equals 2/3 of Saturn's year). 

\begin{figure}
\epsscale{0.7}
\plotone{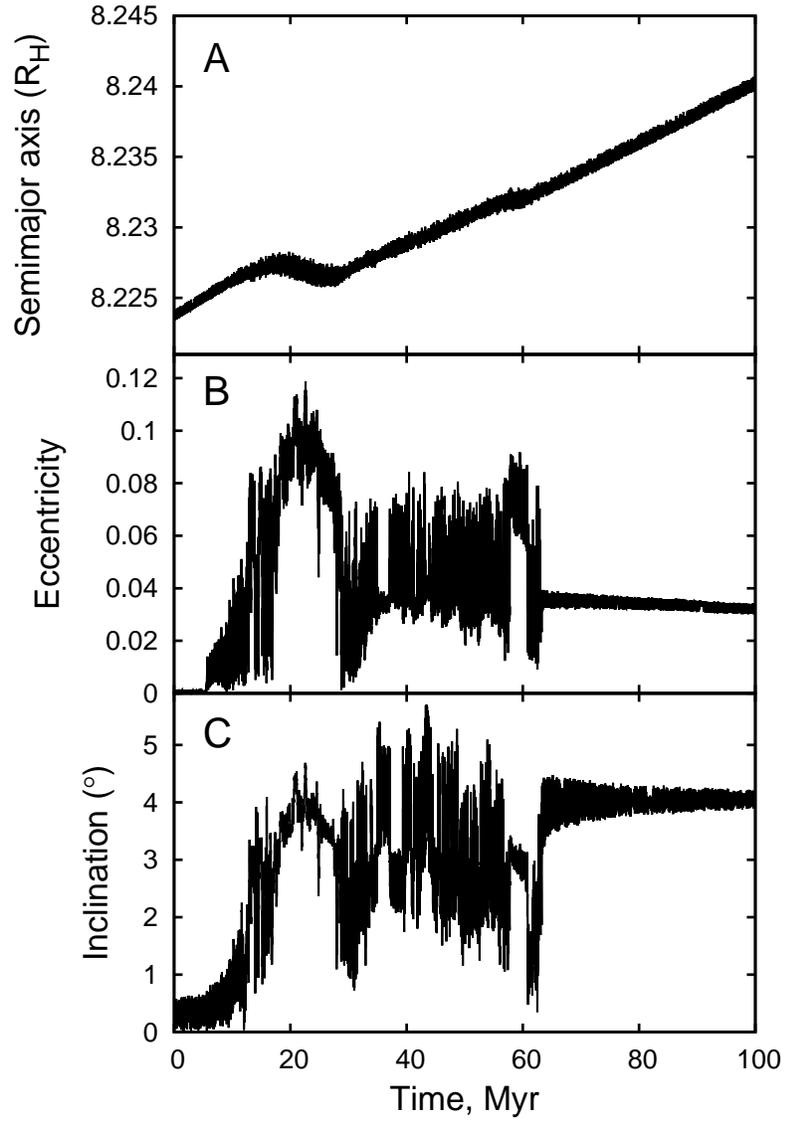}  
\caption{Passage of a moon through the evection resonance with the Sun. We assumed $M=16 \times 10^{20} {\rm \ kg}, Q=100, k_2=0.05$ for the satellite.} 
\label{eve1}
\end{figure}

\begin{figure}
\epsscale{0.7}
\plotone{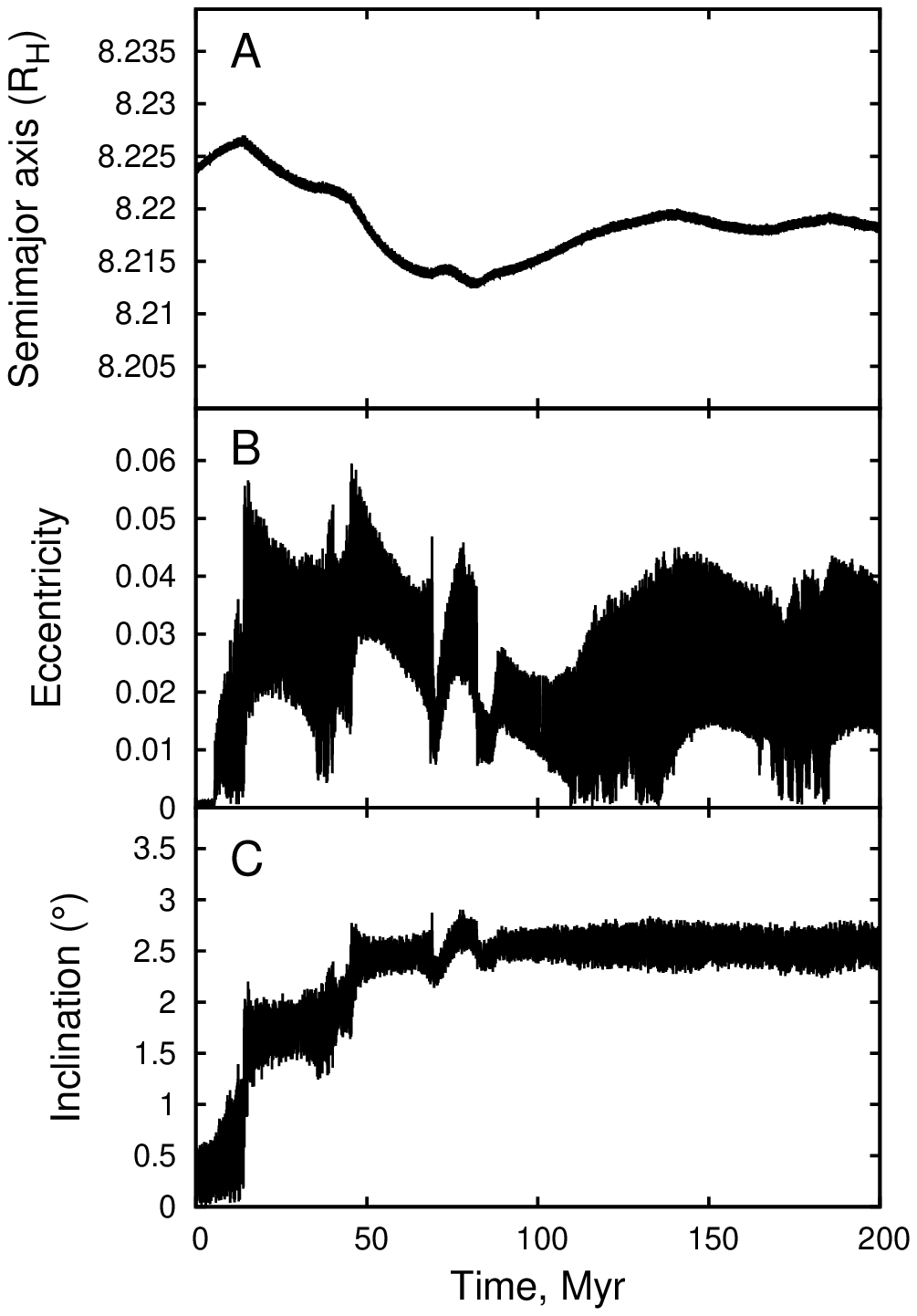}
\caption{Long-lasting slowdown of tidal evolution of an $M=16 \times 10^{20} {\rm \ kg}$ moon due to the evection resonance with the Sun. We assumed $Q=100, k_2=0.5$ for the satellite, which stay constant through the simulation. Unlike in Fig. \ref{eve1}, eccentricity-damping satellite tides are strong enough to prevent the passage through the evection resonance.} 
\label{eve2}
\end{figure}

Figure \ref{eve1} shows a moon encountering the main evection resonance assuming its tidal response stays limited ($M=16 \times 10^{20} {\rm \ kg}, Q=100, k_2=0.05$ for the satellite). The eccentricity of the moon is strongly excited, and so is the inclination. However, since this resonance affects only the eccentricity but not the mean motion of the satellite, and the satellite tides are too weak to reverse the evolution (as we assumed $k_2=0.05$), the satellite eventually passes the resonance without being destabilized (as long as it has no very close neighbors). After crossing the resonance, both eccentricity and inclination are high. While eccentricity can damp, inclination will not, implying that any moon that passed through the evection resonance should have an inclination of several degrees. This is well in excess of the current inclination of Rhea (the only moon beyond the evection resonance), implying that an intact Rhea with these tidal properties could not have formed close to the rings and migrated out to its present location, as suggested by \citet{cha11}. 

If we assume large-scale melting within the moon whose orbit is excited by the evection resonance, its orbital behavior changes. As shown in Fig \ref{eve2} (where we assumed a much larger $k_2=0.5$ for the moon), the evection resonance excites a large eccentricity, which then damps and in the process makes the moon migrate inward. After the eccentricity is damped, the outward migration continues and the cycle repeats. Each next ``bounce" from the resonance will not be identical, as the inclination also evolves. We see no reason why this process cannot continue for 1 Gyr or more, trapping the satellite close to the original location of the evection resonance.

The difference between the scenarios shown in Figs. \ref{eve1} and \ref{eve2} depends only on the internal properties of the satellite. Regardless of the outcome (resonance crossing or long-term stalling of outward tidal evolution), the evection resonance has a potential of exciting satellite orbits to eccentricities of $e=0.1$ or more in very short amounts of time. Before considering the effect of evection on two interacting satellites, we will briefly address some higher-order harmonics of evection.

The evection resonance in the Saturnian system is notably chaotic and tends to affect both eccentricity and inclination. This is in contrast to the evection resonance in the early Earth-Moon system, where it affects only eccentricity, and can lead to stable capture \citep{tou98, cuk12} or a stable near-resonant equilibrium \citep{wis15}. We find that this complex behavior is caused by the evection resonance's overlap with its nearby harmonics, which affect inclination. The principal inclination harmonic has the resonant argument $\lambda_{Sun}-\varpi_{Sun}+\Omega_2-\Omega_{Sun}$, where $\Omega_2$ is the moon's longitude of ascending node, while $\Omega_{Sun}$ is the node of the Sun's apparent orbit measured relative to the moons' Laplace plane (in practice, this angle is equivalent to the equinox of Saturn). This harmonic can be seen as a combination of the annual equation and the principal secular perturbation in inclination \citep{bro61}. This argument's strength varies with Saturn's eccentricity and therefore it cannot lead to stable long-term capture, but it can induce significant variation of inclination, as seen in Figs. \ref{eve1} and \ref{eve2}. In Fig. \ref{eve2}, the inclination is stable after about 100 Myr, indicating that the moon may be in a state where it is affected only by the principal evection resonance.

The strongest among the 2:1 evection harmonics at 6.9 Saturn radii involves not only the perturbed satellite and the Sun, but also Titan. This is due to its resonant argument $2\lambda_{Sun}-\varpi-\varpi_{T}$, where $\varpi$ designates longitude of pericenter and $\lambda$ the mean longitude. This argument combines the basic evection term $2 \lambda_{Sun} - 2 \varpi$ with the moon-Titan secular term $\varpi-\varpi_{T}$ \citep[for a summary of different perturbation terms see][]{bro61}. We find that mid-sized icy moons can get captured in this resonance, and that their further evolution through the resonance increases the eccentricity of Titan. This resonance may seem like an attractive way of generating Titan's eccentricity \citep{cuk13}, but the resonance always breaks due to satellite tides when the smaller moon's eccentricity reaches a few percent. The increase in Titan's eccentricity is always a factor of a few too small to explain the observed value, and this resonance never seems to lead to instability.

The strongest 3:2 evection harmonic at 7.3 Saturn radii has the resonant argument $3\lambda_{Sun}-2 \varpi-\varpi_{Sun}$. This argument can be viewed as a combination of the evection and the annual equation $\lambda_{Sun}-\varpi_{Sun}$ \citep{bro61}. As its argument includes $\varpi_{Sun}$, the 3:2 evection harmonic term's strength is proportional to Saturn's heliocentric eccentricity, which varies by an order of magnitude (0.01-0.09) during a 50,000 year secular cycle \citep{md99}. This means that the resonance is weak during a low point in Saturn's eccentricity, and our simulations show that a capture into this resonance breaks relatively fast, with the moon moving past the resonance with an eccentricity of a few percent and an inclination of a degree or two. We find that this is an unlikely candidate for instability.

\section{Evection Resonance Acting on a Pair of Moons}

In order to produce  a disk, we need a collision between two large moons, probably somewhat similar to Dione and Rhea. As the evection resonance is the most promising source of excitation, we envisage a scenario in which the outer moon is in or near the evection resonance distance, while the inner moon is somewhat closer to Saturn. This configuration can arise either by moons migrating together in a mean-motion resonance, or by the inner moon catching up to the outer one. In general, similar-sized moons would tend to converge on close orbits due to tides, and the real Dione and Rhea (mass ratio 1:2.4) are evolving toward a configuration more compact than their 4:3 mutual resonance. Another possibility is that the outer moon is caught in a cycle of evection resonance ``bounces" (Fig. \ref{eve2}), while the inner moon is catching up.

\begin{figure}
\epsscale{0.7}
\plotone{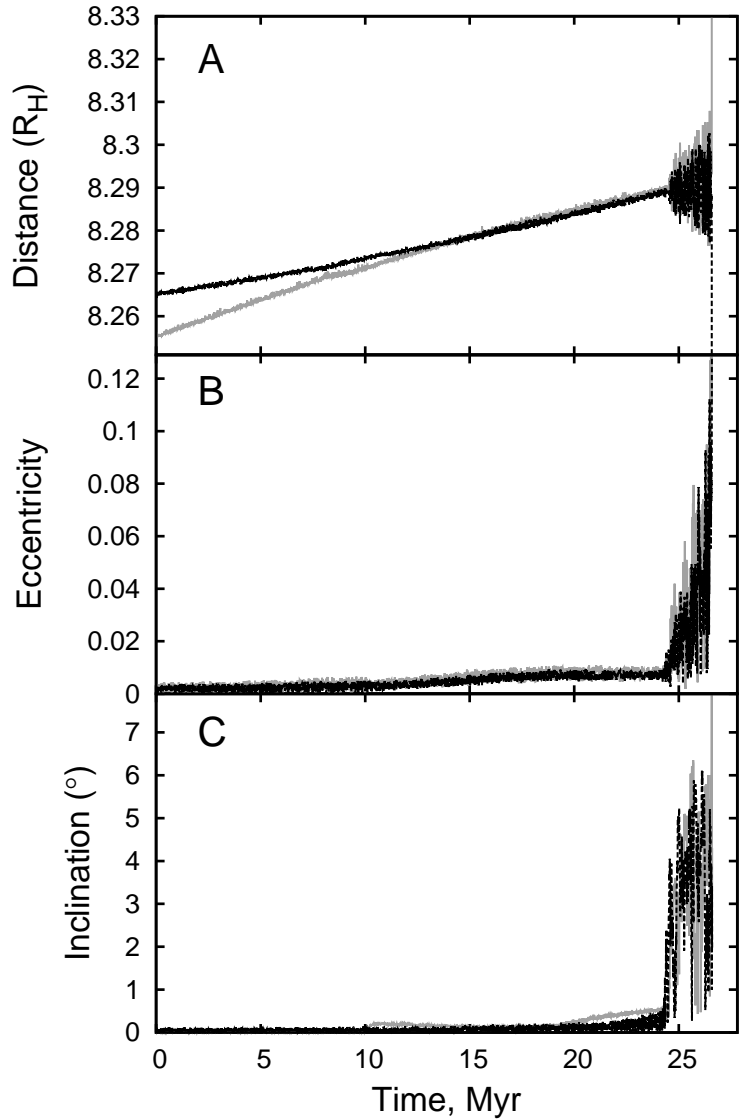}  
\caption{A simulation of two moons resembling Dione and Rhea being destabilized through the evection resonance. Panel A shows the outer moon's semimajor axis (black) and the location of the 4:3 resonance with the inner moon (gray). Panels B and C show the moons' eccentricities and inclinations,  with the inner and outer moon plotted using gray and black lines, respectively.  At first the moons are captured in the 4:3 mean-motion resonance (around 10 Myr into the simulation), which is temporarily stable. About 24 Myr into the simulation the outer moon enters the evection resonance, which excites the orbits of both moons. This excitation leads to orbit crossing and a collision at about 28 Myr.} 
\label{fduo11}
\end{figure}

In Fig. \ref{fduo11} we show an integration (using {\sc simpl}) of convergent migration of a ``proto-Dione" (with a combined mass of Dione and Tethys) and ``proto-Rhea" (with the mass of Rhea) into their mutual 4:3 resonance. Once the outer moon enters the evection resonance, the satellites become dynamically coupled to Saturn's heliocentric orbit and their eccentricities are ultimately excited to large values (e=0.1-0.2), leading to orbit-crossing and a collision. 

\begin{figure}
\epsscale{0.7}
\plotone{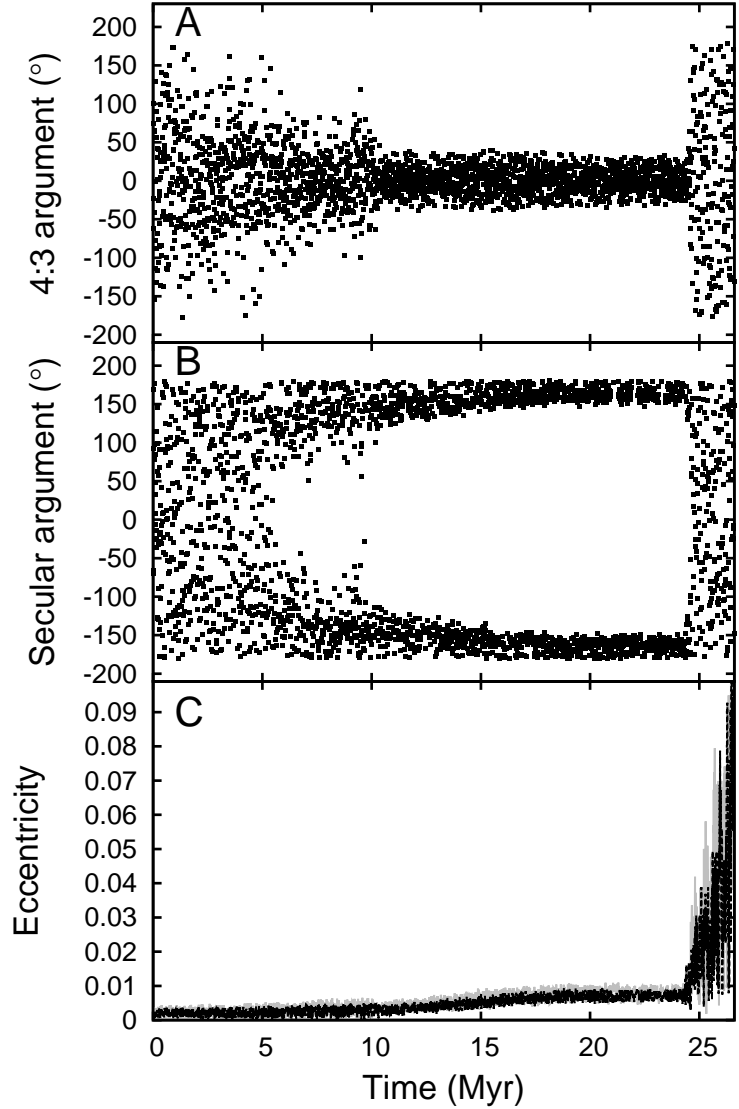}
\caption{The same simulation as shown in Figure \ref{fduo11}. Panel A plots the 4:3 resonant argument $4 \lambda_2 - 3 \lambda_1 - \varpi_1$, Panel B shows the secular resonant argument $\varpi_2 - \varpi_1$, and Panel C shows the  correlated eccentricities of the inner (gray) and outer (black) satellites. $\lambda$ and $\varpi$ designate the mean longitude and the longitude of the pericenter, while subscripts 1 and 2 refer to the inner and the outer satellite, respectively.} 
\label{arg0}
\end{figure}

\begin{figure}
\epsscale{0.7}
\plotone{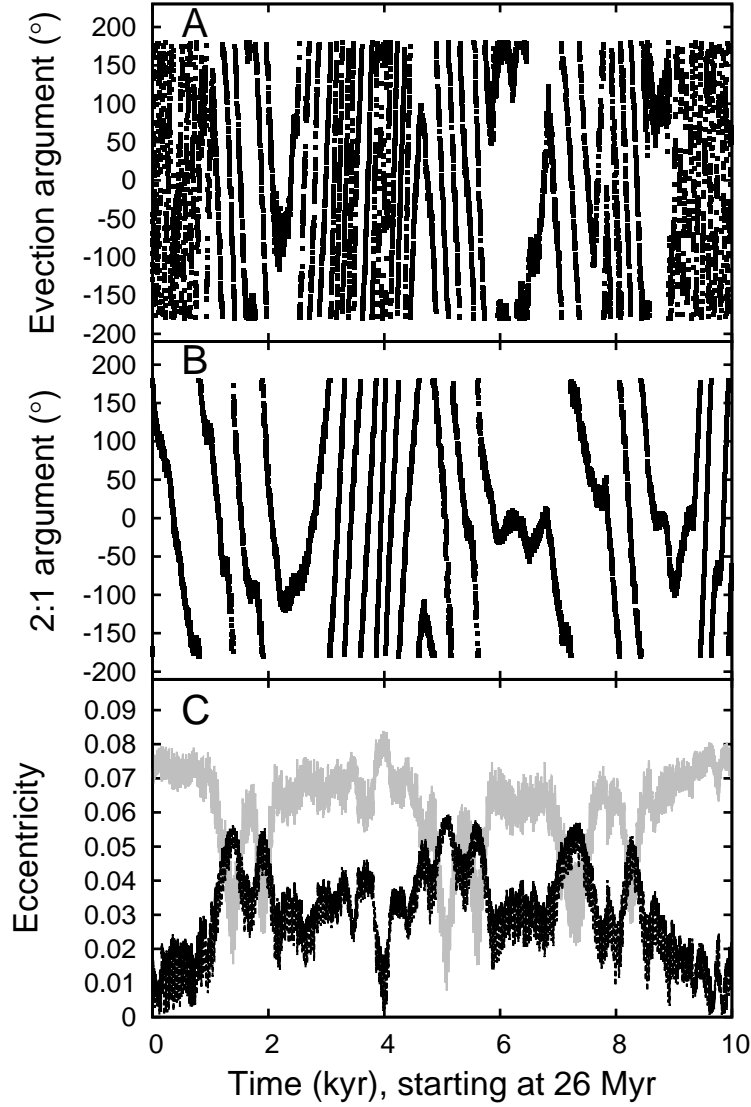}
\caption{The same simulation as shown in Figure \ref{fduo11}, but here we show only 10 kyr of the simulation, starting at 26 Myr. Panel A plots the evection resonance argument $2 \lambda_{Sun} - 2  \varpi_2$, Panel B shows the argument of the 2:1 evection harmonic $ 2 \lambda_{Sun} - \varpi_2 - \varpi_3$, and Panel C shows the eccentricities of the inner (gray) and outer (black) satellite. $\lambda$ and $\varpi$ designate the mean longitude and the longitude of the pericenter, while subscripts 1, 2 and 3 refer to the inner moon, outer moon, and Titan, respectively.} 
\label{arg1}
\end{figure}

\begin{figure}
\epsscale{0.7}
\plotone{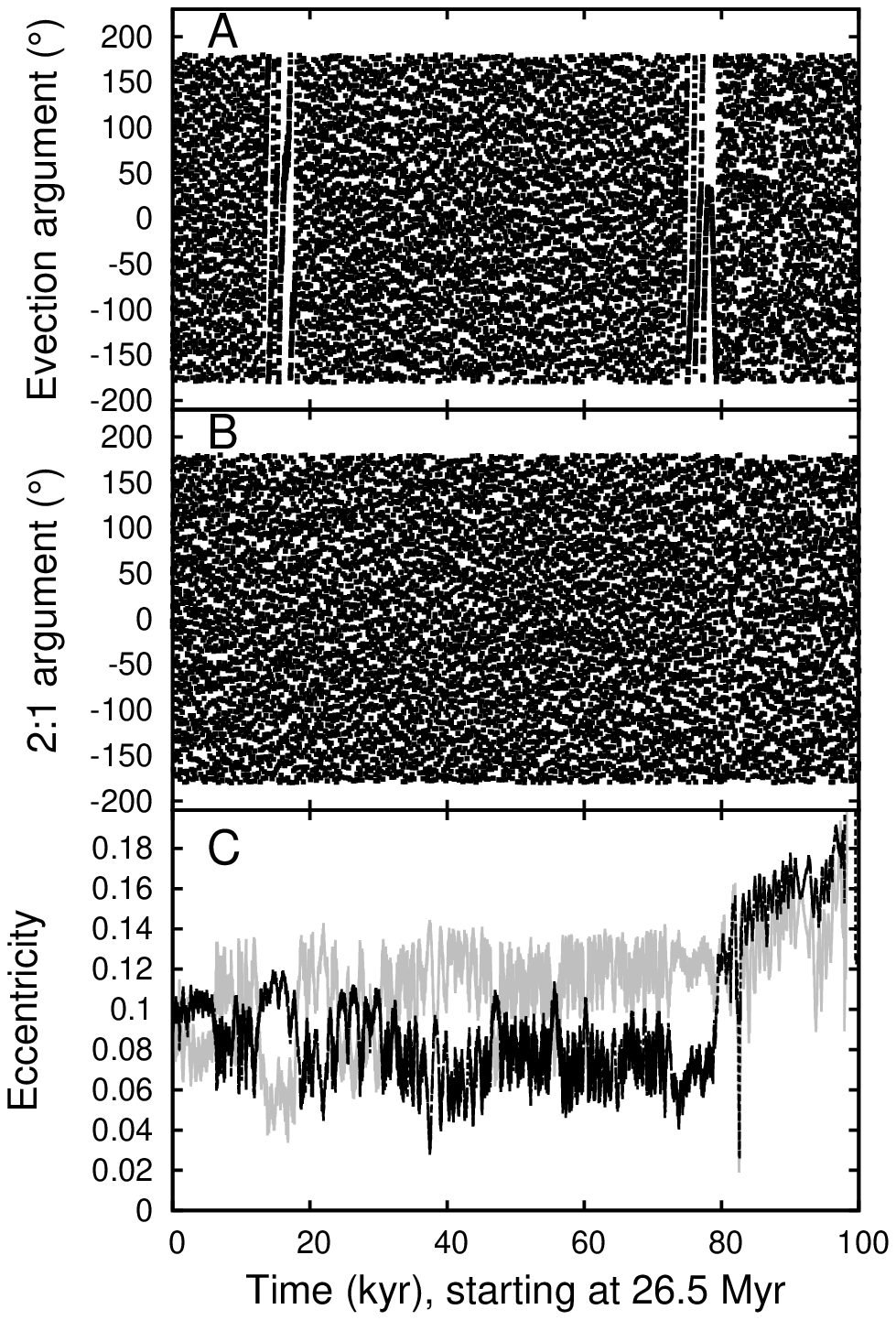}  
\caption{Same simulation as shown in Figure \ref{fduo11}, but here we show only 100 kyr of the simulation, starting at 26.5 Myr. Panel A plots the evection resonance argument $2 \lambda_{Sun} - 2  \varpi_2$, Panel B shows the argument of the 2:1 evection harmonic $ 2 \lambda_{Sun} - \varpi_2 - \varpi_3$, and Panel C shows the eccentricities of the inner (gray) and outer (black) satellite (symbols mean the same as in Fig. \ref{arg1}). The eccentricity boost to the outer moon just before 80 kyr is caused by the evection resonance, while subsequent large kicks are due to close encounters. } 
\label{arg2}
\end{figure}

In this simulation, the 4:3 mean-motion resonance capture happens first, and only then do the moons enter the evection resonance. While the moons are in the 4:3 mean-motion resonance, they are also in the secular resonance $\varpi_2-\varpi_1$ (i.e. apsidal anti-alignment; Fig. \ref{arg0}). Due to the stabilizing effects of the secular resonance the moons are capable of simultaneously being in stable $e_1$ and $e_2$ sub-resonances of the 4:3 resonance, without the chaos usually accompanying resonance overlap (i.e. the arguments $4 \lambda_2 - 3 \lambda_1 - \varpi_1$ and $4 \lambda_2 - 3 \lambda_1 - \varpi_2$ are both librating). This configuration is apparently stable until the outer satellite reaches the evection resonance. 

The next stage of their evolution, starting at about 24 Myr, is chaotic and during this time the moons are not in either 4:3 or (apsidal) secular resonance. Chaos is driven by both the evection resonance proper and its associated harmonics. While the main evection resonance $2 \lambda_{Sun} - 2 \varpi_2$ and its inclination-related harmonic $\lambda_{Sun} - \varpi_{Sun} + \Omega_2 - \Omega_{Sun}$ affect the outer satellite, the 2:1 evection harmonic $\lambda_{Sun} - \varpi_2 - \varpi_T$ affects the inner satellite ($\Omega$ is the longitude of the node, and the subscript T refers to Titan). Fig. \ref{arg1} shows the 10 kyr interval 26 Myr into the simulation, during which both the main evection argument and its 2:1 harmonic reverse the direction of circulation, and this behavior is correlated with the eccentricities of the moons (which continually exchange angular momentum). The presence of the 2:1 evection harmonic at the location of the inner moon is a simple consequence that two moons around an oblate body have an apsidal precession period ratio of about two when they are close to the 4:3 resonance. Since $\dot{\varpi} \propto a^{-7/2}$,  
\begin{equation}
{\dot{\varpi_1} \over \dot{\varpi_2}} = \Bigl({a_2 \over a_1}\Bigr)^{3.5} = \Bigl(\Bigl[{4 \over 3}\Bigr]^{2/3}\Bigr)^{7/2} = 1.96; 
\label{num}
\end{equation}
therefore, if the outer moon's pericenter precesses at a rate close to Saturn's mean motion, the inner moon would naturally precess at about twice that rate (barring other complications), exposing itself to perturbations arising from 2:1 near-resonance with the heliocentric motion. The importance of the 2:1 harmonic depends on the eccentricity of Titan, which was $4 \times 10^{-3}$ in this simulation (i.e. one order of magnitude lower than observed). If Titan efficiently damped any eccentricity it might have had before the last re-arrangement of the system, this resonance may have been much less prominent. In general, we find that the main evection resonance is the most important one for the eventual fate of the system.

The two moons, while having large and rapidly varying eccentricities and inclinations, remain stable for several Myr. Eventually the pair enter a state in which the 2:1 harmonic is less important and the evection resonance dominates. Just before instability, the outer moon gets a major eccentricity boost from the evection resonance (Fig. \ref{arg2}). Soon after that close encounters begin and the two orbits are irreversibly coupled. At about 26.6 Myr, as the orbits become crossing, we pass the simulation from {\sc simpl} to {\sc complex}. We find that the collision typically occurs in about 1000 years. This collision, at about 3 km/s, is just energetic enough to  be erosive and can lead to the formation of a disk \citep{lei12}. 

%\begin{figure}
%\epsscale{0.7}
%\plotone{fduo2.eps}  
%\caption{Instability affecting a pair of satellites, due to the outer moon entering the evection resonance. The gray line plots the elements of the inner moon, while the black line plots those of the outer moon (in panel A, the semimajor axis of the outer moon and the 3:4 resonance with the inner moon are plotted). Unlike in the simulation shown in Fig. 3, the two moons are not in the 4:3 mean motion resonance, but are more widely separated when the instability happens.} 
%\label{duo2}
%\end{figure}

%\begin{figure}
%\epsscale{0.7}
%\plotone{arg3.eps}
%\caption{Same simulation as shown in Figure \ref{duo2}, but here we show only the last Myr of the simulation. Panel A plots the evection resonance argument $2 \lambda_{Sun} - 2  \varpi_2$ and Panel B shows the eccentricities of the inner (gray) and outer (black) satellite (symbols mean the same as in Fig. \ref{arg1}). Note that the excitation of the outer moon's orbit predates that of the inner, and is clearly correlated with episodes of libration in the evection resonance.} 
%\label{arg3}
%\end{figure}

We also achieved instability in a simulation where the moons are initially not in 4:3 resonance, but on orbits converging due to tidal evolution (this time the masses are $M_1=11 \times 10^{20}{\rm \ kg}, M_2=16 \times 10^{20}{\rm \ kg}$). In this simulation we placed the outer moon just interior to the evection resonance, so dramatic effects happen soon. However, given that a moon could be trapped at this distance by the evection resonance for a long time (Fig. \ref{eve2}), it is fully plausible to have the inner moon converge on the outer moon while the latter is trapped by the evection resonance.

%The simulation shown in Fig. \ref{duo2} is easier to understand than the one plotted in Fig. \ref{fduo11}. As the evection resonance dominates the system, only the outer moon experiences strong perturbations from the Sun, which it then passes to the inner moon through the conventional secular interaction. Fig. \ref{arg3} shows the last 1 Myr of this simulation, plotting the outer moon's evection argument alongside the moons' eccentricities. Unlike in Fig. \ref{arg2}, where the two moons' eccentricities are anti-correlated, in this case the eccentricity of the inner moon appears to be driven by that of the outer moon. In turn, the outer moon's eccentricity variation is correlated with episodes of libration in the evection resonance, starting at 17.13 Myr. Particularly notable is the major libration episode starting at 17.6 Myr, during which the outer moon acquires $e_2 > 0.1$, leading to instability. Just like in the case of a single moon, the evection resonance itself is chaotic due to the presence of inclination-related harmonics.

While the collision resulting from the simulation shown in Fig. \ref{fduo11} is marginally energetic enough to disrupt the moons \citep{lei12}, in reality the moons' inclinations may be higher by the time destabilization occurs. A previous residence in either the evection or 4:3 resonance, as well as crossing of any of the other evection harmonics or any other mean motion resonances (with one another or with other satellites) would lead to inclinations of several degrees before the beginning of the last resonance, increasing the likely collision velocity. 

%We cannot state which of the two scenarios of destabilization shown here is more likely to occur. A pair of moons evolving together while locked in the mutual mean-motion resonance is a common outcome of tidal evolution. Also, Fig. \ref{eve2} shows that the evection resonance can arrest tidal evolution of one moon, allowing the other one to catch up. Since the outcome depends sensitively on the masses and starting locations of the previous generation of Saturn's satellites, both situations need to be considered equally likely at this point. In the future, large-scale numerical experiments using many initial configurations may be able to identify the statistically most likely path to instability.

\section{Conclusions}

We used direct numerical simulations to explore a past hypothetical 3:2 mean-motion resonance crossing between Saturn's moons Tethys and Dione, as well as a 5:3 resonance crossing between Dione and Rhea. As the current Tethys/Dione and Dione/Rhea orbital period ratios are just above 2/3 and 3/5 respectively, and tidal recession from Saturn is expected to be fastest for Tethys and slowest for Rhea, these resonances have generally been thought to have been crossed at some point in the past. We find that the 3:2 resonance crossing always leads to excessive excitation of the orbital inclinations of both Tethys and Dione, which is inconsistent with their current orbits. Using the same approach, we conclude that the 5:3 resonance between Dione and Rhea, which closely follows the Tethys-Dione 3:2 resonance in the timeline of tidal evolution, most likely did happen. We find that the 5:3 Dione-Rhea resonance, immediately followed by a previously unknown Tethys-Dione secular resonance, is the most likely source of the current inclinations of Tethys and Rhea. 

We can therefore state that Tethys and Dione evolved tidally by only a modest amount over their lifetimes, which is only about a quarter of the tidal evolution envisaged in \citet{md99}. There are two possible interpretations: either tidal evolution of Saturn's moons has been very slow, or Saturn's mid-sized moons are significantly younger than the Solar System. While both interpretations are consistent with the lack of the past Tethys-Dione resonance, we favor the idea that the moons are young, possibly as young as 100 Myr. Young moons would be consistent with the astrometric analysis of \citet{lai12, lai15}, equilibrium tidal heating of Enceladus \citep{por06, how11, mey07} and the evolved state of the Titan-Hyperion resonance \citep{gre73, cuk13}. Therefore, we propose that Tethys, Dione and Rhea all formed in one large event about 100 Myr ago. The Trojan moons of Tethys and Dione that share their inclinations must have formed even more recently, after their passage through the secular resonance. Mimas, Enceladus and the rings could have formed at the same epoch as Tethys, Dione and Rhea, or could be younger yet. Tidal evolution of Mimas and Enceladus, and their current resonances with Tethys and Dione are a complex subject \citep[][and references therein]{mey08}, and will be addressed in future work.

Our results require most craters on Rhea and the moons interior to its orbit to be produced by planetocentric impactors. Many authors have considered the cratering record of Saturn's mid-sized moons to result from both heliocentric and planetocentric impactors, with the latter being especially important on Mimas and Enceladus \citep{don09, kir10}. Differences in crater size-frequency distributions (SFDs) between moons do not contradict our scenario, as each moon would be primarily cratered by debris on nearby orbits. \citet{bie12, bie15} have shown that differences in surface gravity and impact velocity can produce a wide range of crater SFDs for the moons orbiting a planet. Impact speeds for planetocentric impactors onto the moons will be roughly an order of magnitude smaller than for comets from heliocentric orbit, so larger planetocentric impactors are needed to make a given crater. However, the required impactor sizes remain plausible, as they are still smaller than the crater sizes, even for basins in the Saturn system. For instance, using Keith Holsapple's tool at \url{http://keith.aa.washington.edu/craterdata/scaling/index.htm}, we estimate that a planetocentric impactor with a diameter of 40~km impacting Mimas at 3 km/s could procuce the Herschel basin, which has a diameter of 130~km.  Planetocentric impactors are also fully compatible with more than one cratering episode: the relative youth of Trojan companions and their apparent origin as debris from the larger moons imply at least one intense cratering episode following the resonance crossings described in this paper.

The absence of a past 3:2 resonance between Tethys and Dione also rules out the simplest version of the scenario proposed by \citet{cha11} and \citet{cri12}, in which the moons formed sequentially out of past massive rings, and then migrated outward. That hypothesis implies that Tethys had to ``catch up" with Dione, invariably crossing the 3:2 resonance. Our results do not exclude the possibility that the previous generation of moons did form this way, but were destroyed and re-accreted close to their present positions. Other ultimate origins of the mid-sized satellite system, such as those proposed by \citet{can10} and \citet{asp13}, or a late origin of Titan \citep{ham13}, can certainly be reconciled with a recent episode of re-accretion. Disruptions, followed by formation of a debris disk and re-accretion, erase much of the diagnostic information about the moons' formation, and put few constraints on the ultimate origin of the system. 

%Our result that Tethys and Dione never crossed their mutual 3:2 resonance directly challenges recent proposal that all inner moons sequentially formed close to the rings \citep{cha11, cri12}, as Tethys would need to cross the resonance to reach its present orbit. Note that our conclusions apply only to the (most?) recent episode of satellite formation, and issue of the formation of previous generations of moons is wide open \citep{don91, can10, cha11, asp13}.

Using order-of-magnitude arguments, we conclude that the disk that gave rise to the present mid-sized moons can only be short-lived and must be a product of disruption of a previous generation of mid-sized moons. We explored different configurations of Saturn's past satellite system in order to find a likely source of such an instability. The most likely trigger for a global instability is the evection (semi-secular) resonance between the largest past moon (assumed to be Rhea-sized) and the Sun, located just interior to the present orbit of Rhea. If most of the current icy moons' mass was concentrated in two moons on converging orbits (somewhat similar to Dione and Rhea), with the outer one evolving into the evection resonance, we find that a large scale instability and mutual collisions are a common outcome. We also propose disk-driven migration of small moons (at the edge of the resulting disk) into orbital resonances with Titan as a plausible source of Titan's eccentricity.

There are three possible avenues toward testing the hypothesis of the young Saturnian mid-sized satellite system. While preliminary results appear to support fast migration \citep{lai15}, further work on the satellite astrometry using Cassini data should be able to decisively confirm or falsify the findings of \citet{lai12}. Second, better understanding of the geophysics of Enceladus may be able to clarify if its energy source is indeed tidal and if the dynamical equilibrium hypothesis is appropriate. Third, this hypothesis requires that the majority of craters on Saturn's moons interior to Titan (many of which have heavily cratered surfaces) must be the product of planetocentric impacts, as there would not have been enough time for them to accumulate sufficient numbers of cometary impacts \citep{don09}. We hope that further work on crater and impactor population modeling will be able to put stronger constraints on impactor sources.  

\acknowledgments

M. {\'C}. acknowledges support from NASA's Outer Planets Research Program, awards NNX11AM48G and NNX14AOO38G. We thank an anonymous reviewer for very insightful comments that greatly improved the paper.

%\email{mcuk@seti.org}.

\nopagebreak

\end{document}